\documentclass[acmsmall]{acmart}

\acmJournal{PACMPL}
\acmVolume{1}
\acmNumber{CONF} 
\acmArticle{1}
\acmYear{2018}
\acmMonth{1}
\acmDOI{} 
\startPage{1}

\setcopyright{none}

\usepackage{algorithm}
\usepackage[noend]{algpseudocode}
\usepackage{listings}
\usepackage{mathtools}
\usepackage{multirow}
\usepackage[most]{tcolorbox}
\usepackage{sidecap}
\sidecaptionvpos{figure}{c}

\theoremstyle{definition}

\usepackage{color, soul}
\definecolor{grey}{rgb}{0.9,0.9,0.9}
\usepackage{balance}
\usepackage{textcomp}
\usepackage[T1]{fontenc}

\usepackage{alltt}
\usepackage{microtype}
\usepackage{latexsym}
\usepackage{xspace}
\usepackage{url}
\usepackage{wrapfig}
\usepackage{pdfpages}
\usepackage{alltt}
\usepackage{hyperref}
\usepackage{footnote}
\makesavenoteenv{tabular}
\makesavenoteenv{table}
\usepackage{booktabs}
\usepackage{subcaption}
\usepackage{url}
\usepackage{hhline}
\usepackage{multicol}
\usepackage{makecell}
\usepackage[font={small}]{caption}
\usepackage[inline,shortlabels]{enumitem}
\usepackage{color}
\usepackage[export]{adjustbox}
\usepackage[normalem]{ulem} 
\usepackage{mwe}
\usepackage[scaled=0.75]{beramono}

\hypersetup{colorlinks,
  linkcolor=ACMDarkBlue,
  citecolor=ACMPurple,
  urlcolor=ACMDarkBlue,
  filecolor=ACMDarkBlue}

\xspaceaddexceptions{\}}

\newcommand{\ie}{\emph{i.e.\@}\xspace}

\newcommand{\eg}{\emph{e.g.\@}\xspace}

\newcommand{\emphbf}[1]{\emph{\textbf{#1}\xspace}}
\newcommand{\mypara}[1]{\smallskip\noindent\emphbf{#1.}\xspace}
\newcommand{\mysubsubsection}[1]{\subsubsection{\textbf{#1}}\xspace}

\newboolean{isreview}
\setboolean{isreview}{false}
\ifthenelse{\boolean{isreview}}{
  \newcommand{\replicationpackage}{the supplementary materials}
}{
  \newcommand{\replicationpackage}{
    our replication package\footnote{https://github.com/michaelbjames/copilot-study}}
}

\newboolean{showedits}
\setboolean{showedits}{true} 
\ifthenelse{\boolean{showedits}}
{
	\newcommand{\del}[1]{\textcolor{red}{\sout{#1}}} 
}{
	\newcommand{\del}[1]{} 
	
}

\newboolean{showcomments}
\setboolean{showcomments}{true}
\newcommand{\id}[1]{$-$Id: scgPaper.tex 32478 2010-04-29 09:11:32Z oscar $-$}

\ifthenelse{\boolean{showcomments}}
{\newcommand{\nbc}[3]{
 {\colorbox{#3}{\bfseries\sffamily\scriptsize\textcolor{white}{#1}}}~
 {\textcolor{#3}{\sf\small$\blacktriangleright$\textit{#2}$\blacktriangleleft$}}}
 }
{\newcommand{\nbc}[3]{}
 \renewcommand{\del}[1]{} 
 }

\definecolor{ibcolor}{rgb}{0.4,0.6,0.2}

\definecolor{todocolor}{rgb}{0.9,0.1,0.1}

\definecolor{mjcolor}{rgb}{0.1,0.4,0.1}
\definecolor{shraddhacolor}{rgb}{0.6,0.1,0.3}
\definecolor{nadiacolor}{rgb}{0.3,0.1,0.6}

\newcounter{hypocount}
\definecolor{block-gray}{gray}{0.85}

\newtcolorbox{myquote}[1][]{%
		sharp corners,
    boxrule=0pt,
    boxsep=0pt,
    breakable,
    enhanced jigsaw,
    borderline west={4pt}{0pt}{gray},
    #1,
}

\begin{document}

	\title{Grounded Copilot: How Programmers Interact with Code-Generating Models}

\author{Shraddha Barke}
\authornote{Equal contribution}
\affiliation{
	\institution{UC San Diego}
	\country{USA}  
}
\email{sbarke@eng.ucsd.edu}   

\author{Michael B. James}
\authornotemark[1]
\affiliation{
	\institution{UC San Diego}
	\country{USA}       
}
\email{m3james@eng.ucsd.edu}          

\author{Nadia Polikarpova}
\affiliation{
	\institution{UC San Diego}
	\country{USA}      
}
\email{npolikarpova@eng.ucsd.edu}   

  
\begin{abstract}
Powered by recent advances in code-generating models,
AI assistants like Github Copilot promise to change the face of programming forever.
But what \emph{is} this new face of programming?
We present the first grounded theory analysis of how programmers interact with Copilot,
based on observing 20 participants---%
with a range of prior experience using the assistant---%
as they solve diverse programming tasks across four languages.
Our main finding is that interactions with programming assistants are \emph{bimodal}:
in \emph{acceleration mode},
the programmer knows what to do next and uses Copilot to get there faster;
in \emph{exploration mode},
the programmer is unsure how to proceed
and uses Copilot to explore their options.
Based on our theory,
we provide recommendations for improving the usability of future AI programming assistants.
\end{abstract}


	\maketitle

\section{Introduction}

The dream of an ``AI assistant'' working alongside the programmer
has captured our imagination for several decades now,
giving rise to a rich body of work from both the programming languages \cite{raychev2014code,miltner2019fly,snippy,ni2021recode}
and the machine learning \cite{kalyan2018neural,Xu_2020,guo2021learning} communities.
Thanks to recent breakthroughs in large language models (LLMs)~\cite{Transformer_2017,alphacode}
this dream finally seems within reach.
OpenAI's Codex model~\cite{chen2021evaluating},
which contains 12 billion model parameters
and is trained on 54 million software repositories on GitHub,
is able to correctly solve 30--70\% of novel Python problems,
while DeepMind's AlphaCode~\cite{alphacode} ranked in the top 54.3\%
among 5000 human programmers on the competitive programming platform Codeforces.
With this impressive performance,
large code-generating models are quickly escaping research labs
to power industrial programming assistant tools, such as Github Copilot~\cite{copilot}.

The growing adoption of these tools gives rise to questions
about the nature of AI-assisted programming:
\emph{What kinds of tasks do programmers need assistance with?
How do programmers prefer to communicate their intent to the tool?
How do they validate the generated code to determine its correctness
and how do they cope with errors?}
It is clear that the design of programming assistants should be informed by the answers to these questions,
yet research on these topics is currently scarce.
Specifically, we are aware of only one usability study of Copilot, by \citet{Vaithilingam_2022};
although their work makes several interesting observations about human behavior
(which we discuss in more detail in \autoref{sec:related}),
ultimately it has a narrow goal of measuring whether Copilot helps programmers
in solving stand-alone Python programming tasks.
To complement this study
and to obtain more generalizable insights that can inform the design of future tools,
our work sets out to explore how programmers interact with Copilot
in a broader setting.

\mypara{Our contribution: grounded theory of Copilot-assisted programming}
We approach this goal using the toolbox of \emph{grounded theory} (GT)~\cite{Glaser_Strauss_1967},
a qualitative research technique that has a long history in social sciences,
and has recently been adopted to study phenomena in software engineering~\cite{Stol_Ralph_Fitzgerald_2016}
and programming languages~\cite{Lubin_Chasins_2021}.
GT is designed to build an understanding of a phenomenon from the ground up in a data-driven way.
%
To this end, researchers start from raw data
(such as interview transcripts or videos capturing some behavior)
and tag this data with categories, which classify and explain the data;
in GT parlance, this tagging process is called \emph{qualitative coding}.
Coding and data collection must interleave:
as the researcher gains a better understanding of the phenomenon,
they might design further experiments to collect more data;
and as more data is observed, the set of categories used for coding is refined.

In this paper, we present the first grounded theory
of how users interact with an AI programming assistant---specifically Github Copilot.
To build this theory, we observed 20 participants
as they used Copilot to complete several programming tasks we designed.
Some of the tasks required contributing to an existing codebase,
which we believe more faithfully mimics a realistic software development setting;
the tasks also spanned multiple programming languages---Python, Rust, Haskell, and Java---%
in order to avoid language bias.
We then iterated between coding the participants' interactions with Copilot,
consolidating our observations into a theory,
and adjusting the programming tasks to answer specific question that came up.
The study method is described in detail in \autoref{sec:method}.
%
%
%
%

\mypara{Summary of findings}
The main thesis of our theory (\autoref{sec:theory})
is that user interactions with Copilot can be classified into two modes---%
\emph{acceleration} and \emph{exploration}---%
akin to the two systems of thought in dual-process theories of cognition~\cite{Carlston_2013,MILLI2021104881},
popularized by Daniel Kahneman's ``Thinking, Fast and Slow''~\cite{Kahneman_2011}.
In acceleration mode, the programmer already knows what they want to do next,
and Copilot helps them get there quicker;
interactions in this mode are fast and do not break programmer's flow.
In exploration mode, the programmer is not sure how to proceed
and uses Copilot to explore their options or get a starting point for the solution;
interactions in this mode are slow and deliberate,
and include explicit prompting and more extensive validation.

\autoref{sec:eval} describes two kinds of further analysis of our theory.
First, we performed a quantitative analysis of the data collected during the study,
comparing prompting and validation behaviors across modes,
and quantifying the factors that influence the relative prevalence of each mode. 
Second, to reinforce our findings, we gathered additional data from five livestream videos 
we found on YouTube and Twitch,
and confirmed that the streamers' behavior was consistent with our theory.

Based on our theory,
we provide design recommendations for future programming assistants (\autoref{sec:recs}).
For example, if the tool is aware that the programmer is currently in acceleration mode,
it could avoid breaking their flow
by sticking with only short and high-confidence code suggestions.
On the other hand, to aid exploration,
the IDE could provide better affordances to compare and contrast alternative code suggestions,
or simplify validation of generated code via automated testing or live programming.

\section{Copilot-Assisted Programming, by Example}

Copilot is a programming assistant released by Github in June 2021~\cite{copilot},
and since integrated into several development environments,
including Visual Studio Code, JetBrains and Neovim.
Copilot is powered by the OpenAI Codex family of models~\cite{chen2021evaluating},
which are derived by fine-tuning GPT-3~\cite{brown2020language}
 on publicly available Github repositories.

In the rest of this section,
we present two concrete scenarios of users interacting with Copilot,
which are inspired by real interactions we observed in our study.
The purpose of these scenarios is twofold:
first, to introduce Copilot's UI and capabilities,
and second, to illustrate the two main interaction modes we discovered in the study.

\subsection{Copilot as Intelligent Auto-Completion}
\label{sec:copilot:autocomplete}

\begin{wrapfigure}{r}{.5\textwidth}
    \centering
    \vspace{-10pt}
    \includegraphics[width=.5\textwidth]{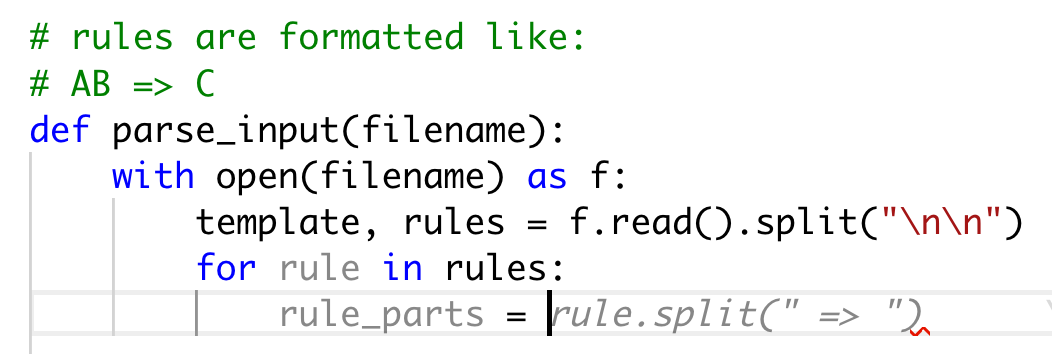}
    \caption{
Copilot's end-of-line suggestion appears at the cursor
without explicit invocation.
The programmer can press \texttt{<tab>} to accept it.
}\label{fig:acc}
\vspace{-10pt}
\end{wrapfigure}

Axel, a confident Python programmer, is solving an Advent of Code ~\cite{aoc} task,
which takes as input a set of rules of the form \texttt{AB => C},
and computes the result of applying these rules to a given input string.
He begins by mentally breaking down the task into small, well-defined subtasks,
the first of which is to parse the rules from the input file into a dictionary.
To accomplish the first subtask,
he starts writing a function \texttt{parse\_input} (\autoref{fig:acc}).
Although Axel has a good idea of what the code of this function should look like,
he thinks Copilot can help him finish it faster and 
save him some keystrokes and mental effort of recalling API function names.
To provide some context for the tool,
he adds a comment before the function definition,
explaining the format of the rules.

As Axel starts writing the function body,
any time he pauses for a second,
Copilot's grayed-out \emph{suggestion} appears at the cursor.
\autoref{fig:acc} shows an example of an \emph{end-of-line suggestion},
which only completes the current line of code.
%
%
In this case, Copilot suggests the correct API function invocation
to split the rule into its left- and right-hand sides.
To come up with this suggestion,
Copilot relies on the \emph{context},
\ie some amount of source file content preceding the cursor,
which can include both code and natural language comments,
as is the case in our example.

Because the suggestion in \autoref{fig:acc} is short and closely matches his expectations,
Axel only takes a fraction of a second to examine and accept it,
without ever leaving his state of flow.
Throughout the implementation of \texttt{parse\_input},
Axel might see a dozen of suggestions,
which he quickly accepts (by pressing \texttt{<tab>}) or rejects (by simply typing on).
Some of them are larger, \emph{multi-line suggestions},
but Axel still seems to be able to dispatch them quickly
by looking for patterns, such as expected control flow and familiar function names.
We liken this kind of interaction with Copilot to the fast \emph{System 1} in dual-process theories of cognition~\cite{Carlston_2013},
which is characterized by quick, automatic, and heuristic decisions.

\subsection{Copilot as an Exploration Tool}
\label{sec:copilot:exploration}

\begin{figure}[h]
    \centering
    \includegraphics[width=0.9\textwidth]{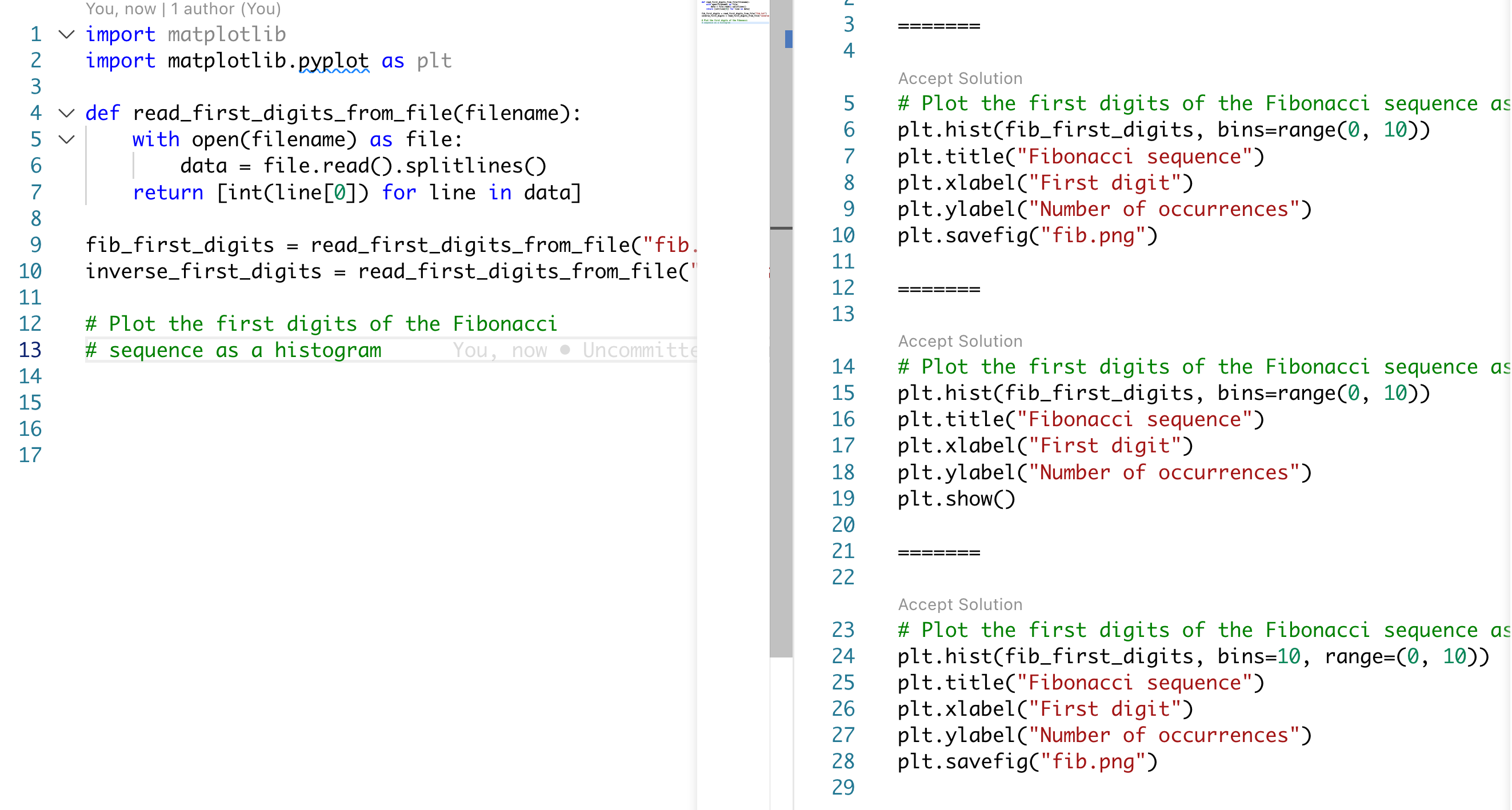}
    \caption{
The user writes an explicit comment prompt (lines 12--13 on the left)
and invokes Copilot's multi-suggestion pane by pressing \texttt{<ctrl> + <enter>}.
The pane, shown on the right, displays up to 10 unique suggestions,
which reflect slightly different ways to make a histogram with \texttt{matplotlib}.
}
    \label{fig:explore}
\end{figure}

\noindent
Emily is new to data science,
and wants to visualize a dataset as a histogram.
While she is familiar with Python,
she is not familiar with the plotting library \texttt{matplotlib}.
As a result, she does not know how to approach this task:
not only which API functions to call,
but also how to decompose the problem
and the right set of abstractions to use.
Emily decides to use Copilot to explore solutions.

Emily explicitly \emph{prompts} Copilot with a natural-language comment,
as shown in lines 12--13 of \autoref{fig:explore}.
Moreover, since she wants to explore multiple options,
she presses \texttt{<ctrl> + <enter>} to bring up the \emph{multi-suggestion pane},
which displays up to 10 unique suggestions in a separate pane
(shown on the right of \autoref{fig:explore}).
Emily carefully inspects the first three suggestions;
since all of them have similar structure
and use common API calls, such as \texttt{plt.hist},
she feels confident that Copilot understands her task well,
and hence the suggestions can be trusted.
She copy-pastes the part of the first suggestion she likes best into her code;
as a side-effect, she gains some understanding of this part of the \texttt{matplotlib} API,
including alternative ways to call \texttt{plt.hist}.
To double-check that the code does what she expects,
Emily runs it and inspects the generated histogram.
This is an example of \emph{validation},
a term we use broadly, to encompass any behavior meant to increase user's confidence 
that the generated code matches their intent.

When faced with an unfamiliar task,
Emily was prepared to put deliberate effort into
writing the prompt,
invoking the multi-suggestion pane,
exploring multiple suggestions to select a suitable snippet,
and finally validating the generated code by running it.
We liken this, second kind of interaction with Copilot to the slow \emph{System 2},
which is responsible for conscious thought and careful, deliberate decision-making.

\section{Method}\label{sec:method}

\begin{table}
	\caption{
Participants overview.
\textsc{PCU}: Prior Copilot Usage.
We show the language(s) used on their task, their usage experience with their
task language (Never, Occasional, Regular, Professional), whether they had used
Copilot prior to the study, their occupation, and what task they worked
on.}
	\label{fig:peopleTable}
	\footnotesize
	\begin{tabular}{|c|l|l|c|l|l|}
		\hline
		\textbf{ID} & \textbf{Language(s)} & \textbf{Language Experience} &
		\textbf{\textsc{PCU}} &\textbf{Occupation} & \textbf{Task}\\
		\hline

		P1 &  Python & Professional &Yes & Professor & Chat Server	  \\
		\hline

		P2 & Rust & Professional & No & PhD Student & Chat Client \\
		\hline

		P3 & Rust & Occasional & No & Professor & Chat Client \\
		\hline

		P4 & Python & Occasional & Yes & Postdoc & Chat Server \\
		\hline

		P5 & Python & Regular & No & Software Engineer  & Chat Client  \\
		\hline

		P6 & Rust & Professional & Yes & PhD Student & Chat Server  \\
		\hline

		P7 & Rust & Professional & No & Software Engineer & Chat Server  \\
		\hline

		P8 & Rust & Professional & No & PhD Student  & Chat Server \\
		\hline

		P9 & Rust\footnote{Participant did not have time to attempt Python section} & Occasional & No & Undergraduate Student & Benford's law \\
		\hline

		P10 & Python & Occasional & No & Undergraduate Student & Chat Client \\
		\hline

		P11 & Rust+Python & Professional + Professional & Yes & Cybersecurity Developer & Benford's law \\
		\hline

		P12 & Rust+Python & Professional + Occasional & Yes & Software Engineer & Benford's law \\
		\hline

		P13 & Rust+Python & Regular + Occasional & Yes & PhD Student & Benford's law \\
		\hline

		P14 & Python & Professional & No & PhD Student & Advent of Code \\
		\hline

		P15 & Python & Professional & Yes & PhD Student & Advent of Code \\
		\hline

		P16 & Haskell & Professional & No & PhD Student & Advent of Code \\
		\hline

		P17 & Rust & Professional & Yes & Founder & Advent of Code \\
		\hline

		P18 & Java & Occasional & No & PhD Student & Advent of Code \\
		\hline

		P19 & Python & Occasional & No & PhD Student & Advent of Code \\
		\hline

		P20 & Haskell & Occasional & Yes & PhD Student & Advent of Code \\
		\hline
	\end{tabular}
\end{table}

\mypara{Participants}
We developed our theory through a user study with 20 participants
(15 from academia and 5 from industry).
We recruited these participants through
personal contacts, Twitter, and Reddit. 
Nine of the participants had used Copilot to varying degrees prior to the study.
Participants were not paid, but those without access to Copilot were provided
access to the technical preview for continued use after the study concluded.
\autoref{fig:peopleTable} lists relevant information about each participant.
We asked each participant to select a statement best describing their level of
experience with
possible target languages, with options ranging from ``I have never used
Python'', to ``I use Python professionally'' (from least-to-most, used in
\autoref{fig:peopleTable}: Never, Occasional, Regular, and Professional).
%
We screened out participants who had never used the target language.
We choose a qualitative self-assignment of experience as other common metrics,
such as years-of-experience, can be misleading.
For example, a professor having used Rust occasionally over eight years
is arguably less experienced than
as a software engineer using Rust all day for a year.


\mypara{User protocol}
To study participants using Copilot, we gave them a programming task to attempt with
Copilot's help.
Over the course of an hour a participant was given a small training task to
familiarize them with Copilot's various usage models (\ie code completion, natural language
prompt, and the multi-suggestion pane).
During the core task---about 20-40 minutes---a participant was asked to talk
through
their interactions with Copilot.
They were encouraged to work Copilot into their usual workflow,
but they were not required to use Copilot.
After the task, the interviewer asked them questions through a semi-structured
interview; these questions as well as the tasks are available in
our supplementary package.
The entire session was recorded and transcribed to use as data in our grounded
theory.

\mypara{Grounded Theory Process}
Grounded Theory (GT) takes qualitative data and produces a theory in an
iterative process, first pioneered by~\cite{Glaser_Strauss_1967}.
As opposed to evaluating fixed, a priori hypotheses, a study using the GT
methodology seeks to generate new hypotheses in an overarching theory
developed without prior theoretical knowledge on the topic.
A researcher produces this theory by constantly interleaving data collection and
data analysis.
%
GT has diversified into three primary styles over the past half-century.
We follow the framework laid out by Strauss and Corbin~\cite{Strauss_Corbin_1990},
commonly called \emph{Straussian Grounded
Theory}~\cite{Stol_Ralph_Fitzgerald_2016}.
We describe our process below.

%
%
%
We began our study with the blank slate question: 
``How do programmers interact with Copilot?''
Our bimodal theory of acceleration and exploration was not yet formed.
%
%
During each session, we took notes to guide our semi-structured
interview.
After each session, we tagged any portion of the recording relevant to Copilot
with a note.
We took into account what the participant said, what they did, and their body
language.
For example, we initially tagged an instance where P2 was carefully examining and
highlighting part of a large Copilot
suggestion as ``validating sub-expression''.
Tagging the data in this way 
is called \emph{(qualitative) coding}; 
and doing so without a set of predefined codes is called \emph{open coding} in Straussian GT.
%
%
%
%
The first two authors coded the first two videos together, 
to agree on a coding-style, 
but later data were coded by one and discussed by both.

%
By the end of the eighth session, we began to see patterns emerging in our data.
We noticed two distinct patterns in our codes which eventually crystallized
into our acceleration and exploration modes.
%
%
During this phase of analysis, we aggregated our codes to understand the
\emph{conditions} when a
participant would enter acceleration or exploration, and the \emph{strategies} a
participant deployed in that mode.
For example, we realized that if a programer can decompose a problem, then they
often ended up in acceleration (details in
\autoref{sec:acceleration:breakingDownTasks}).
Once in this acceleration mode,
programmers would validate a
suggestion by a kind of visual ``pattern matching''
(details in \autoref{sec:acceleration:patternMatching}).
This process of aggregating and analyzing our codes form the \emph{axial coding}
phase of GT.

%
After the eighth session, we created a new task to specifically test our
emerging theory.
This process of testing aspects of a theory-in-progress is known in GT as
\emph{theoretical sampling}.
After gathering sufficient data on that third task, 
we created a fourth task to investigate one final aspect of our theory
(validation of Copilot's suggestions).
In the second half of the study, we linked together our codes and notes into the
final bimodal theory we present, in what Straussian GT calls \emph{selective
coding}.
At the 20th participant, we could fit all existing data into our theory and no
new data surprised us.
Having reached this point of \emph{theoretical saturation}, we concluded our GT
study.

\mypara{Tasks}
The list of all four tasks and their descriptions can be found in
\autoref{fig:tasktable}.
The full task templates we provided to participants are available in
\replicationpackage.

Our tasks evolved over the course of the study.
We started with the ``Chat Server'' and ``Chat Client'' pair of tasks, meant to
emulate working on a complex project, with a shared library and specialized APIs.
These two initial tasks required contributing to an existing codebase we created,
which implements a secure chat application.
The first task, Chat Server, asked participants to implement the server backend,
focusing on its ``business logic''.
We provided most of the networking code,
and the participant's task was to implement
the log-in, chat, and chat-command functionality (\eg \texttt{/quit} to quit).
The complementary task Chat Client focused on the client side of the chat application.
Here, we provided no networking code
so the participant had to figure out how to use the often unfamiliar socket API.
We also required using a custom cryptographic API we implemented,
in order to ensure that some part of the API was unfamiliar both to the
participant and to Copilot.
%

To investigate the acceleration and exploration modes further, we created the
``Benford's Law''\footnote{
Benford's Law
says that in natural-looking datasets,
the leading digit of any datum is likely to be small.
It is useful as a signal for finding fraudulent data.} task.
This task had two parts, to separately investigate the acceleration and exploration modes
we found.
In the first half, the participant implements an efficient Fibonacci sequence
generator.
We believed that all participants would be familiar with the algorithm, 
and hence would accelerate through this half of the task,
allowing us to more deeply characterize the acceleration mode.
%
%
%
In the second half, they plotted this sequence and another
($\frac{1}{2},\frac{1}{3},...,\frac{1}{180}$ as \texttt{float}s) using
\texttt{matplotlib};
this sub-task is used as the example in \autoref{sec:copilot:exploration}.
Our participants were not confident users of the plotting library's API,
so they needed to turn to some external resource to complete the task.
This half stressed the code exploration part of our theory.
In addition, our Benford's Law task asked participants to complete the first half in Rust and
the second half in Python.
This division gave us within-participant information on
how different languages impact Copilot usage.

Our fourth task was a string manipulation problem inspired by the 2021 edition of Advent of Code
(this task is used as the example in \autoref{sec:copilot:autocomplete}).
%
%
%
We wanted to collect more data about how programmers validate suggestions from Copilot, 
and this task was a good fit
because it comes with a test case and a very precise description,
and also has two independent sub-tasks,
so it provided several options for checking solutions at different levels of granularity.
%
The data we collected rounded out our hypotheses about validation
(\autoref{sec:acceleration:patternMatching}, \autoref{sec:exploration:validation}).

\begin{table}
	\caption{The four programming tasks used in our study and what high level purpose they served.}
	\label{fig:tasktable}
\footnotesize
\begin{tabular}{|p{2cm}|p{2cm}|p{4.25cm}|p{4cm}|}
\hline
\textbf{Task} & \textbf{Language(s)} &\textbf{Description} &\textbf{Purpose} \\
\hline

Chat Server & Python/Rust &
Implement core ``business logic'' of a chat application,
involving a small state machine.
&
Investigate how Copilot aids in interpreting and implementing a human-language
specification. \\
\hline

Chat Client & Python/Rust &
Implement networking code for a chat application,
using a custom cryptographic API and
standard but often unfamiliar socket API.
&
Probe how Copilot can aid with a custom API. \\
\hline

Benford's Law & Rust \& Python &
Use Rust to generate two sequences---%
the Fibonacci sequence and reciprocals of sequential natural numbers;
then plot these sequences using Python's \texttt{matplotlib}.
&
Collect data on a straightforward task (acceleration) and on an unfamiliar task
(exploration).
\\
\hline

Advent of Code& Python/Rust/ Haskell/Java &
Implement a string manipulation task from a programming competition.
&
How will programmers test their Copilot-assisted code?
\\
\hline
\end{tabular}
\end{table}
\section{Theory}\label{sec:theory}

Through our grounded theory analysis, we identified two main modes of developer
interactions with Copilot:
\emph{acceleration} and \emph{exploration}.
In acceleration mode, a programmer uses Copilot to \emph{execute} their planned code actions,
by completing a logical unit of code or a comment.
Acceleration works within user's sense of flow.
For example, recall how in \autoref{sec:copilot:autocomplete}
Axel accepted Copilot's suggestion of \texttt{rule.split(" => ")},
knowing it was what he wanted to type anyways.
This is a characteristic example of acceleration,
where Copilot was helping him program faster.

In exploration mode, a programmer relies on Copilot to help them \emph{plan} their code actions.
A programmer may use Copilot to assist with unfamiliar syntax,
to look up the appropriate API,
or to discover the right algorithm.
In \autoref{sec:copilot:exploration},
when Emily was searching for the right set of \texttt{matplotlib} calls,
she was considering alternatives,
gaining confidence in the API,
and simply trying to learn how to finish her task.
All of these intentions are part of the exploration mode when using Copilot.
We found that programmers alternate between these two modes as they complete their task,
fluidly switching from one mode to the other.

In this section, we systematize our observation of each mode:
\emph{acceleration} (\autoref{sec:acceleration}) and
\emph{exploration} (\autoref{sec:exploration}).
For each mode, we start with identifying the \emph{conditions}
that lead the participant to end up in that mode,
and then proceed to describe common \emph{strategies} (\ie behavioral patterns) 
we observed in that mode.
Each numbered subsection (\eg \autoref{sec:acceleration:breakFlow}) is a hypothesis deriving from our
top-level bimodal theory.
Each named paragraph heading is an aspect of that hypothesis.

\subsection{Acceleration}
\label{sec:acceleration}


Acceleration is characterized by the programmer being ``in the zone'' and
actively ``driving'' the development,
while occasionally relying on Copilot to complete their thought process.
%
%
A programmer will often accept a Copilot suggestion without much comment and
keep on going without losing focus.
In this interaction mode, programmers tend to think of Copilot as an intelligent
autocomplete that just needs to complete their line of thought.
This idea was well put by P13 who said:
\begin{myquote}
``I think of Copilot as an intelligent autocomplete... I already have the line
of code in mind and I just want to see if it can do it, type it out faster than
I can.''
\end{myquote}
\noindent
P15 added to this,
calling Copilot ``more or less an advanced autocomplete''.
%

\mysubsubsection{Programmers use acceleration after decomposing the task}
\label{sec:acceleration:breakingDownTasks}
We found that the main causal condition
for a participant to end up in acceleration mode
is being able to decompose the programming task into \emph{microtasks}.
%
We define a microtask to be a participant-defined task
with a well-understood and well-defined job.
%
For example, when P16 was working on the Advent of Code task,
they created two separate microtasks to parse the input
and to compute the output.
%
Because they understood these microtasks well,
they wrote a type signature and used descriptive names for each of them;
as a result, Copilot was adept at completing these microtasks for them.
%
%
Another illustrative example is our Benford's Law task,
which was explicitly designed to have a familiar and an unfamiliar subtask.
In the first subtask, participants were asked to implement a fast Fibonacci function.
All four participants were familiar with the Fibonacci sequence
and knew how to make it efficient.
%
As a result, all of them were able to use Copilot to accelerate through this familiar microtask.
P14 explicitly noted:
\begin{myquote}
``I think Copilot would be more helpful in cases where there are a lot of
tedious subtasks which requires less of thinking and more of just coding.''
\end{myquote}

We observed that language expertise or familiarity with Copilot 
seem to play less of a role in determining whether a participant would engage in acceleration,
compared to their understanding of the \emph{algorithm} for solving the task.
For example, P4 was not very comfortable with Python,
but they knew what needed to be done in their task algorithmically,
and so were able to break it down into microtasks,
leading to acceleration.
That said, we do observe that language experts and prior Copilot users
spend a larger proportion of their total interaction time in acceleration mode;
we present quantitative data supporting this observation in \autoref{sec:eval}.



\mysubsubsection{Programmers focus on small logical units}
\label{sec:acceleration:logicalUnits}
Participants who interacted with Copilot in acceleration mode
would frequently accept end-of-line suggestions.
These were often function calls or argument completions.
For example, when P1 wanted to send a
message to a client connection object in the Chat Client task, they typed
\texttt{client\_conn.se} and
immediately accepted Copilot's suggestion \texttt{client\_conn.send\_message()}.
This behavior was seen across all the four tasks when participants were in
acceleration mode.
For a microtask of parsing file input, P15 wanted to spilt the data based on
spaces so they typed \texttt{data = x.} to which Copilot correctly suggested
\texttt{data = x.split("") for x in data}.
Participants would happily accept these end-of-line completions with reactions
like ``Yes that's what I wanted!'' and ``Thank you Copilot!''

When a programmer is focused on a logical unit of code,
they want suggestions \emph{only for that unit}.
When they are writing a print statement, 
they prefer to get a suggestion to the end of the statement.
When writing a snippet to message all connected clients,
they might instead prefer an entire \texttt{for} loop,
\emph{but not more}. 
%
For example, at one point P8 was focused on a single call to the \texttt{startswith} function,
but Copilot suggested a large piece of code;
%
P8 reacted with ``that's way more than what I needed!'' and went on to delete
everything but the first line \texttt{if msg.startswith('/')}.
%

The size of a logical unit differs based on the language and context.
In an imperative language, this is most often a line of code.
However, in a functional language like Haskell,
logical units appear to be smaller.
P16 said that 
``in Haskell it just needs to suggest less. 
[It should] give me the next function I'm going to compose and not the whole composition chain.'' 
%

%



\mysubsubsection{Long suggestions break flow}
\label{sec:acceleration:breakFlow}
In acceleration mode, long, multi-line suggestions are at best dismissed out of hand
and at worst distract the programmer away from their flow.

Upon getting a 16-line suggestion and after just four seconds of review P6 uttered:
``Oh God, no. Absolutely not''.
When P6 got other large suggestions, they would exclaim, ``Stop it!'',
and continue to program as before.
This participant also made use of the \texttt{<esc>} key binding to actively
dismiss a suggestion they did not care for.


%
On the other hand, many programmers felt ``compelled to read the [suggested] code'' (P16)
and noted that reading long suggestions would often break their flow.
As P1 puts it:
\begin{myquote}
  ``When I'm writing, I already have in mind the full line and it's just a matter
	of transmitting to my fingertips, and to the keyboard. But when I have those
	mid-line suggestions and those suggestions are not just until the end of
	line, but actually a few more lines, that breaks my flow of typing. So
	instead of writing the full line, I have to stop, look at the code, think
	whether I want this or not.''
\end{myquote}
\noindent
This sentiment was echoed by multiple participants:
P11 was ``distracted by everything Copilot was throwing at [them]'';
P7 was ``lost in the sauce'' after analyzing a long suggestion;
P17 felt ``discombobulated'', and others (P8, P11) made similar comments. 
P16 put it eloquently:
\begin{myquote}
``I was about to write the code and I knew what I wanted to write. But now I'm
	sitting here, seeing if somehow Copilot came up with something better than the
	person who's been writing Haskell for five years, I don't know why am I
	giving it the time of day.''
\end{myquote}

Such distractions cause some programmers to give up on the tool entirely: 
P1, P6, and P15 all had Copilot disabled prior to the study---%
having had access for several months---%
and they all cited distractions from the always-on suggestions as a factor.

\mysubsubsection{Programmers validate suggestions by ``pattern matching''}
\label{sec:acceleration:patternMatching}
%
%
%
%
In order to quickly recognize whether a suggestion is worthwhile, participants
looked for the presence of certain keywords or control structures.
The keywords included function calls or variable names that
they expected should be part of the solution.
P1 explicitly stated that the presence or absence of certain keywords would
determine whether the suggestion was worth considering.

Most other programmers who commented on how they validated suggestions in acceleration mode
mentioned control structures (P4, P17, P19).
%
%
P4, for instance, immediately rejected an iterative suggestion because they strongly
preferred a recursive implementation.
%
%
%
On one occasion, Copilot suggested code to P6 when they already had an idea
of what shape that code should take;
they described their validation process in this instance as follows:
\begin{myquote}
``I have a picture in mind and that picture ranges from syntactic or textual
	features---like a literal shape in words---to semantic about the kind of
	methods that are being invoked, the order in which they should be invoked,
	and so on. When I see a suggestion, the closer that suggestion is to the
	mental image I hold in my head, the more likely I am to trust it.''
\end{myquote}

%
%
%
These participants appear to first see and understand control-flow features
before understanding data or logic flow features.
This is consistent with previous findings dating back to FORTRAN and COBOL
programming \cite{Pennington_1987}, where programmers briefly shown small code
snippets could best answer questions about control flow compared to data- or
logic-flow.
%

%
%



\mysubsubsection{Programmers are reluctant to accept or repair suggestions}
\label{sec:accelertion:reluctant}
Participants in acceleration mode end up quickly rejecting suggestions that
don't have the right patterns.
Suggestions that are almost-correct were accepted if a small repair was
obvious to the participant.
%
P1 accepted a small inline suggestion which had a call to \texttt{handshake()}
function, checked if it existed, and since it did not,
they made a minor modification, changing the function name to \texttt{do\_dh\_handshake()}.
%
The entire accept-validate-repair sequence seemed to occur without interrupting their state of flow.
P1, P4 would often accept similar-looking function names but
double check if they actually existed:
\begin{myquote}
``Each time it uses something else from the context, I usually double check,
like in this case it was very similar so I could have been fooled, and each
time
this happens it reinforces the need to check everything just to see if it
has the proper names.''
\end{myquote}

Although programmers tend to dismiss code that does not match their expectations,
sometimes Copilot's suggestion makes them aware of a corner case
they have not yet considered.
%
%
%
%
P4 saw Copilot write an inequality check while working on the Chat Server task,
and they said that they
``probably wouldn't have remembered on their first run through to check that [clients] are
distinct''.
Both P6 and P8, working in Rust on the Chat Server,
%
noticed that Copilot used a partial function \texttt{.unwrap()}.
%
%
%
%
When asked about this, P8 said:
\begin{myquote}
	``Copilot suggested code to handle it in one case and now I'm going to change
	it around to handle the other case as well.''
\end{myquote}
%


\subsection{Exploration}
\label{sec:exploration}

%
%

In the previous section we focused on the use of Copilot
when the programmer has a good idea for how to approach the task.
But what if they do not?
In that case they might use Copilot to help them get started,
suggest potentially useful structure and API calls,
or explore alternative solutions.
All of these behaviors fit under what we call exploration mode.
Exploration is characterized by the programmer letting Copilot "drive", as opposed to acceleration, where the programmer is the driver.
%
%
In the rest of this section,
we first describe the conditions that lead programmers to enter exploration mode,
and then we characterize the common behaviors in that mode.

\mysubsubsection{Programmers explore when faced with novel tasks or unexpected behavior}
\label{sec:exploration:novel}

Recall that most often the programmer ended up in acceleration mode
once they had successfully decomposed the programming task into a sequence of steps
(\autoref{sec:acceleration:breakingDownTasks});
dually, when the programmer was uncertain how to break down the task,
they would often use Copilot for code exploration.
%
%
P4 said:
\begin{myquote}
``Copilot feels useful for doing novel tasks that I don't necessarily know how to
do.
It is easier to jump in and get started with the task''.
\end{myquote}
Not knowing where to start was one of two primary ways we observed participants
begin an exploration phase of their study.
The other way participants (P11, P13, P14) began exploration was when they hit
some code that does not work as expected,
regardless of the code's provenance.
They would try a variety of prompting and validation strategies to attempt to
fix their bug.
%
%
%
%
%
%
%
%
%
%
%

\mysubsubsection{Programmers explore when they trust the model}
\label{sec:exploration:trust}

A participant's level of confidence and excitement about code-generating models
was highly correlated with whether and to which extent they would engage in exploration.
During the training task, Copilot produced a large, correct suggestion for P18;
they exclaimed, ``I'm not gonna be a developer, I'm gonna be a
guy who comments!''
This level of excitement was shared among many of our participants early in the
task, like P7 saying, ``it's so exciting to see it write [code] for you!''.
Those participants who were excited about Copilot would often 
let the tool drive
before even attempting to solve the task themselves.

Sometimes, such excessive enthusiasm would get in the way of actually completing a task.
%
For example, P10 made the least progress compared to others on the same task;
in our post-study interview, they admitted that they were, 
``a little too reliant on Copilot'':
%
\begin{myquote}
	``I was trying to get Copilot to do it for me, maybe I should have given
	smaller tasks to Copilot and done the rest myself instead of depending
	entirely on Copilot.''
\end{myquote}
This overoptimism is characteristic of the misunderstanding users often have
with program synthesizers.
P9 and P10 were both hitting the \emph{user-synthesizer gap}, which separates what
the user expects a program synthesizer to be capable of, and what the synthesizer
can actually do \cite{snippy}.

\mysubsubsection{Programmers explicitly prompt Copilot with comments}
\label{sec:exploration:comments}
Nearly every participant (P2, P3, P4, P5, P7, P8, P10, P11, P12, P13, P14, P17,
P18, P19, P20) wrote at least one natural language comment as a prompt to
Copilot, specifically for an exploratory task.

%
%
\mypara{Programmers prefer comment prompts in exploration}
Programmers felt that natural language prompts in the form of comments offered a greater
level of control than code prompts (P17).
P2 told us that,
``writing a couple of lines [of comments] is a lot easier than writing code.''
This feeling of being more in control was echoed by P5 who said:
\begin{myquote}
	``I think that the natural language prompt is more cohesive because it's
	interruptive to be typing out something and then for Copilot to guess what
	you're thinking with that small pseudocode. It's nice to have a comment
	that you've written about your mental model and then going to the next line
	and seeing what Copilot thinks of that.''
\end{myquote}

%
%
%
%
%
%

%
%
\mypara{Programmers write more and different comments when using Copilot}
Participants seem to distinguish between comments made for themselves and Copilot.
In the words of P6,
``The kind of comments I would write to Copilot are not the kind of comments I would use to document my code.''
P2, P3, P5, P12, and P19 all told us that the majority of their comments 
were explicitly meant for Copilot.
P7 was the sole exception: 
they wrote comments to jot down their design ideas saying, 
``I'm writing this not so much to inform Copilot but just to organize my own thoughts'';
they added that being able to prompt Copilot using those comments was a nice
side effect.

Participants were willing to invest more time interacting with Copilot via comment prompts in exploration mode.
They would add detailed information in the comments in the hope that
Copilot would have enough context to generate good suggestions (P2, P3).
They would rewrite comments with more relevant information if the suggestions
did not match their expectations, engaging in a conversation with Copilot.
P2 and P6 wished they had a ``community guide'' (P2) on how to write comments
so that Copilot could better understand their intent.

Further, in our interviews, multiple people described their usual commenting workflow
as post-hoc: they add comments after completing code.
Hence, the participants were willing to change their commenting workflow to
get the benefits of Copilot.

\mypara{Programmers frequently remove comments after completing an interaction with Copilot}
%
Many participants (P3, P4, P7, and P8) would repeatedly delete comments that
were meant for Copilot.
P19 said that cleaning up comments written for Copilot is essential:
\begin{myquote}
	``I wrote this comment to convert String to array just for Copilot, I would never
	leave this here because it's just obvious what it's doing. 
	%
	[\ldots]
	These comments aren't adding value to the code.
	I think you also have to do like a comment cleanup after using Copilot.''
\end{myquote}

\mysubsubsection{Programmers are willing to explore multiple suggestions}
\label{sec:exploration:multipleSuggestions}

In exploration mode, we often saw participants spend significant time
foraging through Copilot's suggestions in a way largely unseen during acceleration.
This included using the \emph{multi-suggestion pane},
both for its primary intended purpose---selecting a single suggestion out of many---%
and for more creative purposes,
such as cherry-picking snippets from multiple suggestions,
API search,
and gauging Copilot's confidence in a code pattern.


Participants tend to use the multi-suggestion pane when faced with
an exploratory task 
(P2, P4, P5, P7, P10, P12--20).
They would either write a comment prompt or a code prompt before invoking
the multi-suggestion pane.
This enabled participants to explore alternate ways to
complete their task while also providing an explicit way to invoke Copilot.
P10, P15, P19 preferred the multi-suggestion pane over
getting suggestions inline in all cases.
P15 said:
\begin{myquote}
	``I prefer multiple suggestions over inline because sometimes the first
	solution is not what I want so if I have something to choose from, it makes
	my life easier.''
\end{myquote}
Some only occasionally got value from the multi-suggestion pane.
P6 said that:
\begin{myquote}
	``If I think there's a range of possible ways to do a task and I want Copilot
	to show me a bunch of them I can see how this could be useful.''
\end{myquote}
Similar to P6, P14 and P17 preferred the multi-suggestion pane only while exploring
code as it showed them more options.
Yet others turned to the multi-suggestion pane when Copilot's always-on
suggestions failed to meet their needs.
%
%


\mypara{Programmers cherry-pick code from multiple suggestions}
Participants took part of a
solution from the multi-suggestion pane or combined code from
different solutions in the pane.
P2, P3, P4, P5, P18 often accepted only interesting sub-snippets from the
multi-suggestion pane.
For example, P18 forgot the syntax for declaring a new \texttt{Hashmap} in Java, 
and while Copilot suggested a bunch of formatting code around the suggestion, 
P18 only copied the line that performed the declaration.
P2 went ahead to combine interesting parts from more than one suggestion stating:

\begin{myquote}
	``I mostly just do a deep dive on the first one it shows me, and if that
	differs from my expectation, for example when it wasn't directly
	invoking the handshake function, I specifically look for other suggestions
	that are like the first one but do that other thing correctly.''
\end{myquote}

\mypara{Programmers use the multi-suggestion pane in lieu of StackOverflow}
%
When programmers do not know the immediate next steps in their workflow, they often
write a comment to Copilot and invoke the multi-suggestion pane.
This workflow is similar to how programmers already use online forums like StackOverflow:
they are unsure about the implementation details but they can describe their goal.
In fact, P12 mentioned that they were mostly using the multi-suggestion pane as a
search engine during exploration.
P4 often used Copilot for purely syntactic searches, for example, to find
the \texttt{x in xs} syntax in Python.
P15 cemented this further:
\begin{myquote}
	``what would have been a StackOverflow search, Copilot pretty
	much gave that to me.''
\end{myquote}
%

Participants emphasized that the multi-suggestion pane helped them use unfamiliar APIs,
even if they did not gain a deep understanding of these APIs.
P5 explains:
\begin{myquote}
	"It definitely helped me understand how best to use the API.
	I feel like my actual understanding of [the socket or crypto library] is not better
	but I was able to use them effectively."
\end{myquote}

\mypara{Programmers use the multi-suggestion pane to gauge Copilot's confidence}
Participants assigned a higher confidence to Copilot's suggestions if a particular
pattern or API call appeared repeatedly in the multi-suggestion pane.
Participants seemed to think that repetition implied Copilot was more confident about the
suggestion.
For example, P5 consulted Copilot's multi-suggestion pane when they were trying to use the
unfamiliar socket library in Python.
After looking through several suggestions, and seeing that they all called the same method,
they accepted the first inline suggestion.
When asked how confident they felt about it, P5 said:
\begin{myquote}
	``I'm pretty confident. I haven't used this socket library, but it seems
	Copilot has seen this pattern enough that, this is what I want.''
\end{myquote}
P4 had a similar experience but with Python syntax:
they checked the multi-suggestion pane to reach a sense of consensus with Copilot
on how to use the \texttt{del} keyword in Python.
%
%

\mypara{Programmers suffer from cognitive overload due to multi-suggestion pane}
P1, P4, P6, P7 and P13 did not like the
multi-suggestion pane popping up in a separate window stating that it added
to their cognitive load.
%
P4 said that they would prefer a modeless (inline) interaction, and P6 stated:
\begin{myquote}
	``Seeing the code in context of where it's going to be was way more valuable
	than seeing it in a separate pane where I have to draw all these additional
	connections.''
\end{myquote}

P13 spent a considerable amount of time skimming
and trying to differentiate the code suggestions in the multi-suggestion pane,
prompting them to make the following feature request:
\begin{myquote}
``It might be nice if it could highlight what it's doing or which parts
are different, just something that gives me clues as to why I should pick
one over the other.''
\end{myquote}
%

\mypara{Programmers suffer from an anchoring bias when looking through
multiple suggestions}
The anchoring bias influences behavior based on the first piece of information
received.
We observed participants believe that suggestions were ranked and that the top
suggestion \emph{must} be closest to their intent (P18).
%
This was also evident through P2's behavior who would inspect the first suggestion more deeply then skim through the rest.

\mysubsubsection{Programmers validate suggestions explicitly}
\label{sec:exploration:validation}

Programmers would validate Copilot's suggestions more carefully
in exploration mode as compared to acceleration.
%
%
Their validation strategies included code \emph{examination},
code \emph{execution} (or testing),
relying on IDE-integrated \emph{static analysis} (\eg a type checker),
and looking up \emph{documentation}.
We look at these techniques in detail.

\mypara{Examination}
Unlike acceleration mode,
where participants quickly triage code suggestions by "pattern matching",
exploration mode is characterized by carefully examining the details of Copilot-generated code.
For example, P19 said that they would ``always check [the code] line by line'',
and P5 mentioned that their role seemed to have shifted from
being a programmer to being a code reviewer:
``It's nice to have code to review instead of write''.
Participants found it important to cross-check Copilot's suggestions just as they would
do for code from an external resource.
When asked how much they trusted Copilot's suggestions, P14 said:
\begin{myquote}
``I consider it as a result I would obtain from a web search.
It's not official
documentation, it's something that needs my examination...if it works it
works'' 
\end{myquote}


\mypara{Execution}
Code execution was common---occurring in every task by at least one participant---%
although not as common as examination.
%
In case of the server and client task, participants P3 and P7 would
frequently run their code by connecting the client to server and
checking if it has the expected behavior.
For the Benford's law task, P11 wrote test cases in Rust using
\texttt{assert\_eq} to check whether the Fibonacci function suggested by Copilot
was correct.
All participants in the Advent of Code task ran their code to check whether they
parsed the input file correctly.

In addition to executing the entire program,
some participants used a Read-Eval-Print-Loop (REPL) as a scratchpad to validate
code suggestions (P14, P16, P19).
P16 used the Haskell REPL throughout the study to validate the results of
subtasks. 
Copilot suggested an \texttt{adjacents} function that
takes a string a pairs adjacent characters together.
%
P16 validated the correctness of this function by running it
on toy input \texttt{adjacents "helloworld"}.
%

\mypara{Static analysis}
In typed languages like Rust, the type checker or another static analyzer 
frequently replaced validation by execution.
For example, P17 did not run their code even once for the Advent of Code task,
despite the task being designed to encourage testing.
They reasoned that the \texttt{rust-analyzer}\footnote{https://github.com/rust-lang/rust-analyzer} tool---%
which compiles and reports type errors in the background---%
took away the need to explicitly compile and execute the code.

\begin{myquote}
	``In Rust I don't [explicitly] compile often just because I feel like the
	there's a lot of
	the type system and being able to reason about state better because
	mutability is demarcated a lot. But if this were in Python, I would be
	checking a lot by running in a REPL.''
\end{myquote}

P7 thought it was ``cool how you can see the suggestion and then rely on the
type checker to find the problems.''
In general, most participants using statically typed languages relied on IDE
support to help them validate code.
P6, P8 and P17 relied on Rust analyzer and P6 had this to say:
\begin{myquote}
	``%
	I rely on the Rust compiler to check that
	I'm not doing anything incorrect. The nice part about being a statically
	typed language is you can catch all that at compile time so I just rely on
	Rust analyzer to do most of the heavy lifting for me.''
\end{myquote}

%
%
\mypara{Documentation}
Lastly, consulting documentation was another common strategy to explicitly validate
code from Copilot.
As an example,
P11 was trying to plot a histogram in \texttt{matplotlib}, but was
unsure of the correct arguments for the
\texttt{plt.hist} function.
They accepted a couple of Copilot's suggestions but explicitly went to validate
the suggested arguments by reading the documentation within the IDE.
Participant P17, who never executed their Rust code,
would instead hover over the variables and function names to access API documentation within the IDE.
Participants that did not have documentation built into their IDE would turn to
a web resource.
For example, P14 accepted Copilot's suggestion for parsing file input in the Advent of Code task,
and then validated the functionality of \texttt{splitlines} by crosschecking
with the official Python documentation.
P11 also used Google for crosschecking whether the Fibonacci sequence suggested
by Copilot was accurate.

%
%
%
%
%
%
%
%
%
%
%
%

\mysubsubsection{Programmers are willing to accept and edit}
\label{sec:exploration:acceptAndEdit}

Unlike acceleration mode, where participants were quick to dismiss a suggestion
that didn't match their expectations,
during exploration, they seemed to prefer deleting or editing code
rather than writing code from scratch.
When a participant saw a segment of code that they felt they were likely to need
in the future, they would hang on to it (P2, P3, P4, P6, P8).
P2 was exploring code for one stage of writing a chat server when they saw code
needed for a later stage and said: ``I'm keeping [that] for later''.
During their exploration, they accepted a 40 line block of code to add:
\begin{myquote}
	``Eh, I'm just going to accept this. It's close enough to what I want that I
	can modify it.''
\end{myquote}
\noindent
P3 said: ``I wanna see what
[Copilot] gives me, then I'll edit them away''.
Some participants were able to complete most of their task by accepting a large
block of code and then slowly breaking it down.
P7 accepted a large block of code early on and iteratively repaired it into the code they needed.
P5 had a similar experience and said,
``It's nice to have code to review instead of write''.

Commonly, participants sought a suggestion from Copilot only to keep the control structure.
As a representative example,
P8 was writing a message-handling function in Rust,
when Copilot produced a  15-line suggestion, 
containing a \texttt{match} statement and the logic of its branches.
After examination, P8 accepted the suggestion but quickly
deleted the content of the branches, retaining only the structure of the
\texttt{match}.
We saw this many times with P1, P2, P11, P17, P18 as well.
P17 said:
\begin{myquote}
	``If I'm in a mode where I want to rip apart a solution and use it as a
	template then I can look at the multi-suggestion pane and select
	whichever suits my needs.''
\end{myquote}

\mypara{Copilot-generated code is harder to debug}
On the flip side,
participants found it more difficult to spot an error in code generated by Copilot.
%
For example, P13 had to rely on Copilot to interface with \texttt{matplotlib};
%
when they noticed undesired behavior in that code, they said:
\begin{myquote}
	``I don't see the error immediately and unfortunately because this is
	generated, I don't understand it as well as I feel like I would've if I had
	written it.
	I find reading code that I didn't write to be a lot more difficult than
	reading code that I did write, so if there's any chance that Copilot is
	going to get it wrong, I'd rather just get it wrong myself because at least
	that way I understand what's going on much better.''
\end{myquote}
\noindent
We observed a similar effect with P9, 
who could not complete their task due to subtly incorrect code suggested by Copilot.
Copilot's suggestion opened a file in read-only mode,
causing the program to fail when attempting to write.
%
%
%
P9 was not able to understand and localize the error,
instead spending a long time trying to add more code 
to perform an unrelated file flush operation.

\section{Additional Analysis}\label{sec:eval}

\begin{figure}
	\includegraphics[max size={0.95\textwidth}{\textheight}]{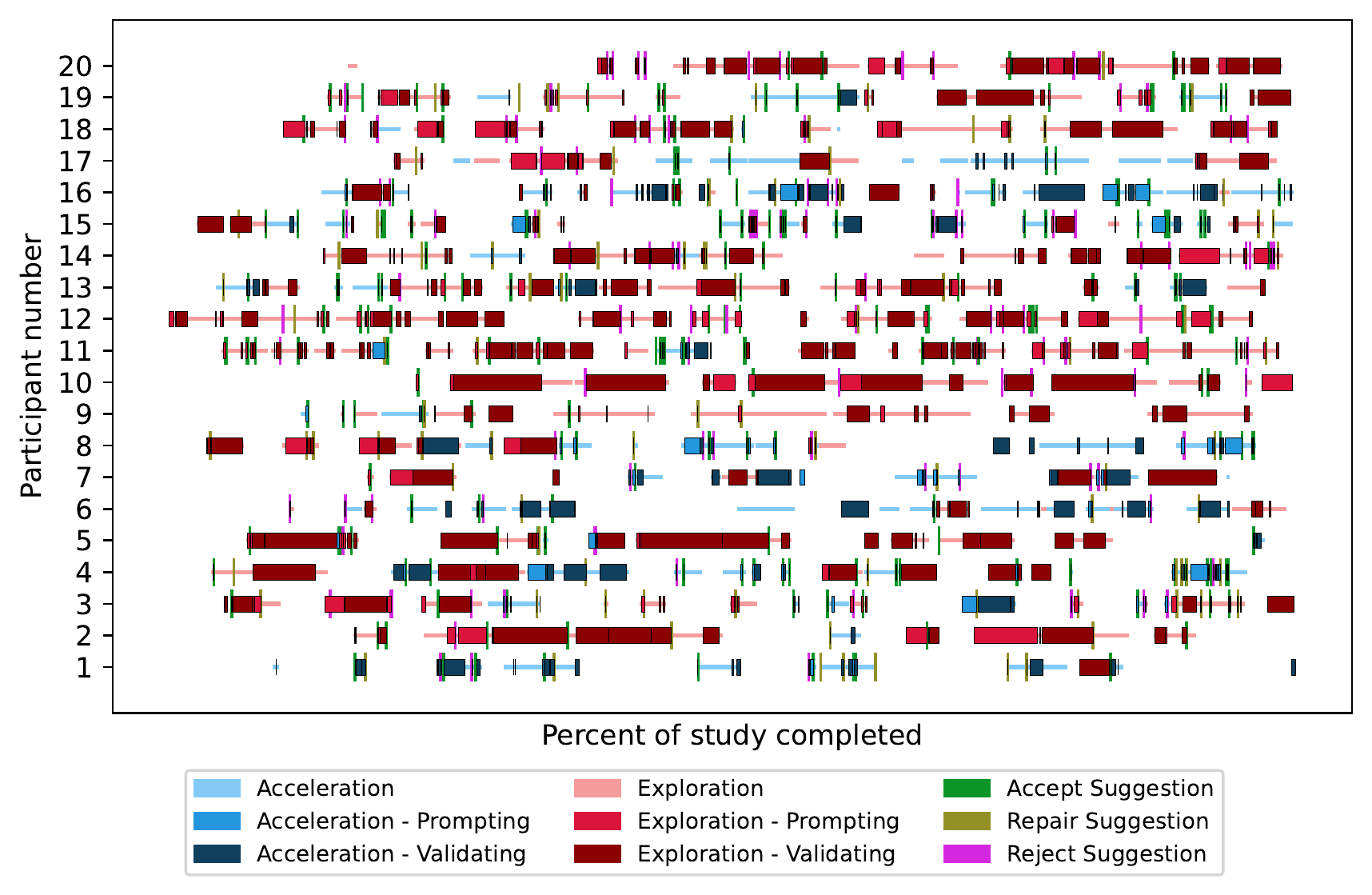}
	\caption{Timeline of observed activities in each interaction mode for the 20 study participants. The qualitative codes include different prompting strategies, validation strategies and outcomes of Copilot's suggestions (accept, reject or repair)}
	\label{fig:codeline}
\end{figure}

In this section, we first provide quantitative evidence to support the findings from our grounded theory analysis.
We then present the results of a qualitative analysis on five livestream videos to provide additional evidence that further supports our theory.

\subsection{Quantitative Analysis}
\label{sec:quant:analysis}
At the end of our grounded theory analysis, we
closed our codebook and re-coded all videos with a fixed set of codes 
that emerged to be most noteworthy.
\autoref{fig:codeline} represents this codeline of the different activities we
observed in each of the two interaction modes. The activities include prompting
strategies, validation strategies, and the outcomes of Copilot's suggestions \ie
whether the participant accepts, rejects, or edits the suggestion.
%
We then performed a quantitative analysis on this codeline to investigate the following questions:

\vbox{%
\begin{enumerate}[label=(\bfseries Q\arabic*)]
	\item What factors influence the time spent in each of the two interaction modes?
	\item What are the prompting strategies used to invoke Copilot in the two interaction modes?
	\item How do the validation strategies differ across the two interaction modes and by task?
\end{enumerate}
}

\begin{figure}[]
	\centering
	\begin{subfigure}{.5\textwidth}
		\centering
		\includegraphics[width=\textwidth]{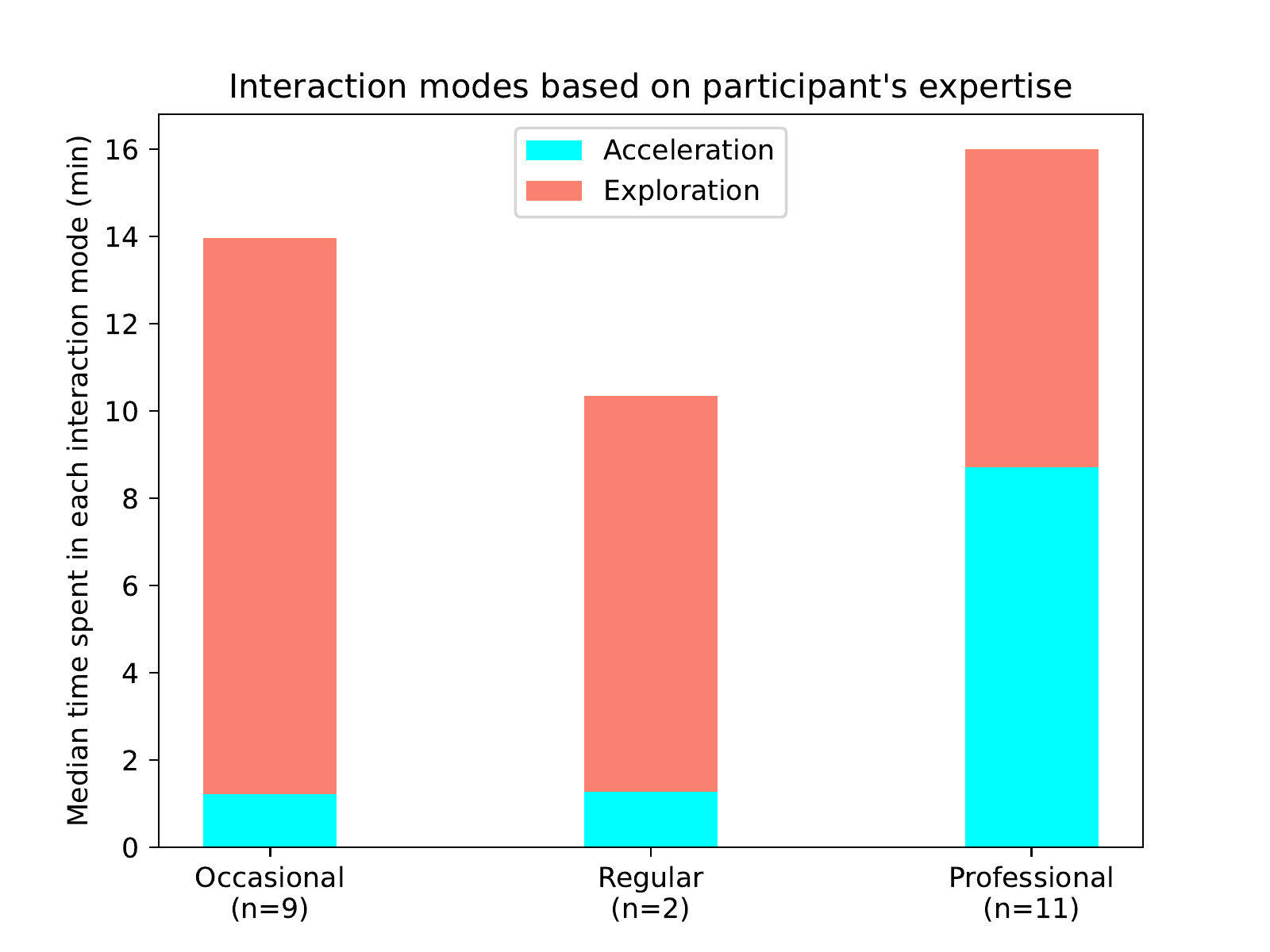}
		\caption{Grouped by language expertise.}
		\label{fig:modes_expertise}
	\end{subfigure}%
	\begin{subfigure}{.5\textwidth}
	\centering
	\includegraphics[width=\textwidth]{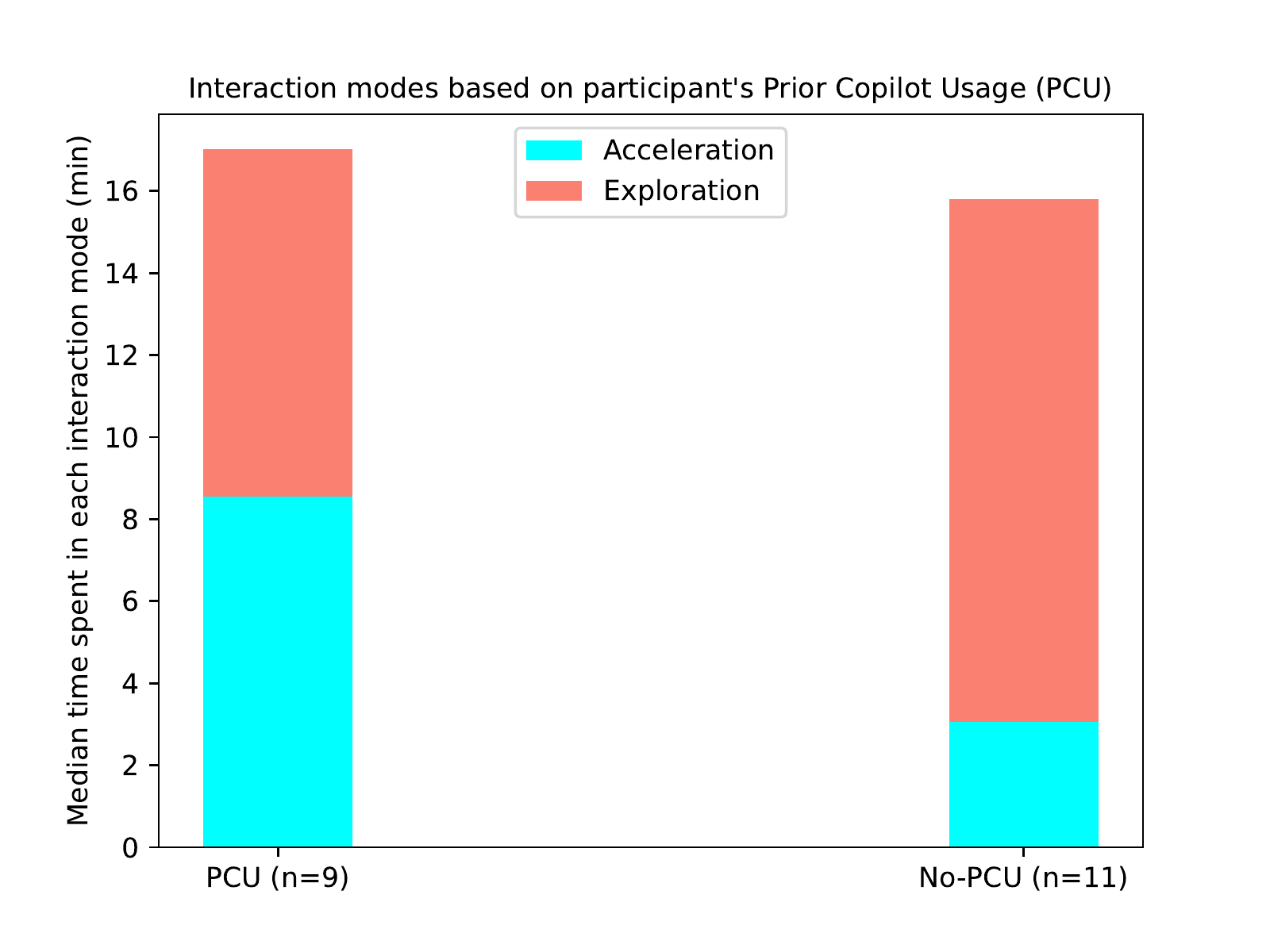}
	\caption{Grouped by prior Copilot usage.}
	\label{fig:modes_pcu}
\end{subfigure}%
  \caption{Median time spent in acceleration vs exploration mode for different participant groups.}
\end{figure}

\mysubsubsection{Time spent in interaction modes}
\label{sec:5.3}
The total amount of study time spent by all participants
interacting with Copilot in exploration mode (248.6 minutes) 
is more than twice that in acceleration mode (104.7 minutes).
This is not surprising,
since exploration is the ``slow \emph{System 2}'' mode,
where each interaction takes longer.
%
At the same time, the ratio of time spent in the two modes is not constant across participants.
Below, we investigate which factors influence this ratio,
including language expertise, prior Copilot usage, the nature of the task, and the programming language.

\mypara{Language expertise}
 \autoref{fig:modes_expertise} shows the median time spent in two modes split by the participant's language expertise.
 We can clearly see that professional participants with the most language expertise
 spend more time accelerating than the other two groups.
 %
 This is not surprising, since they are more likely to already know how to solve the task in the given language.

\mypara{Prior Copilot usage}
We can see in \autoref{fig:modes_pcu} that the total interaction time is
roughly the same for participants with and without prior Copilot usage.
Given roughly the same overall time, 
prior users spend less time exploring (and more time accelerating) than novice users.
%
We attribute this difference to the effect we observed in \autoref{sec:exploration:trust},
where novice users have higher expectations of Copilot's ability to solve high-level exploratory tasks.

\begin{figure}
	\begin{minipage}{0.5\textwidth}
		\centering
		\includegraphics[width=\textwidth]{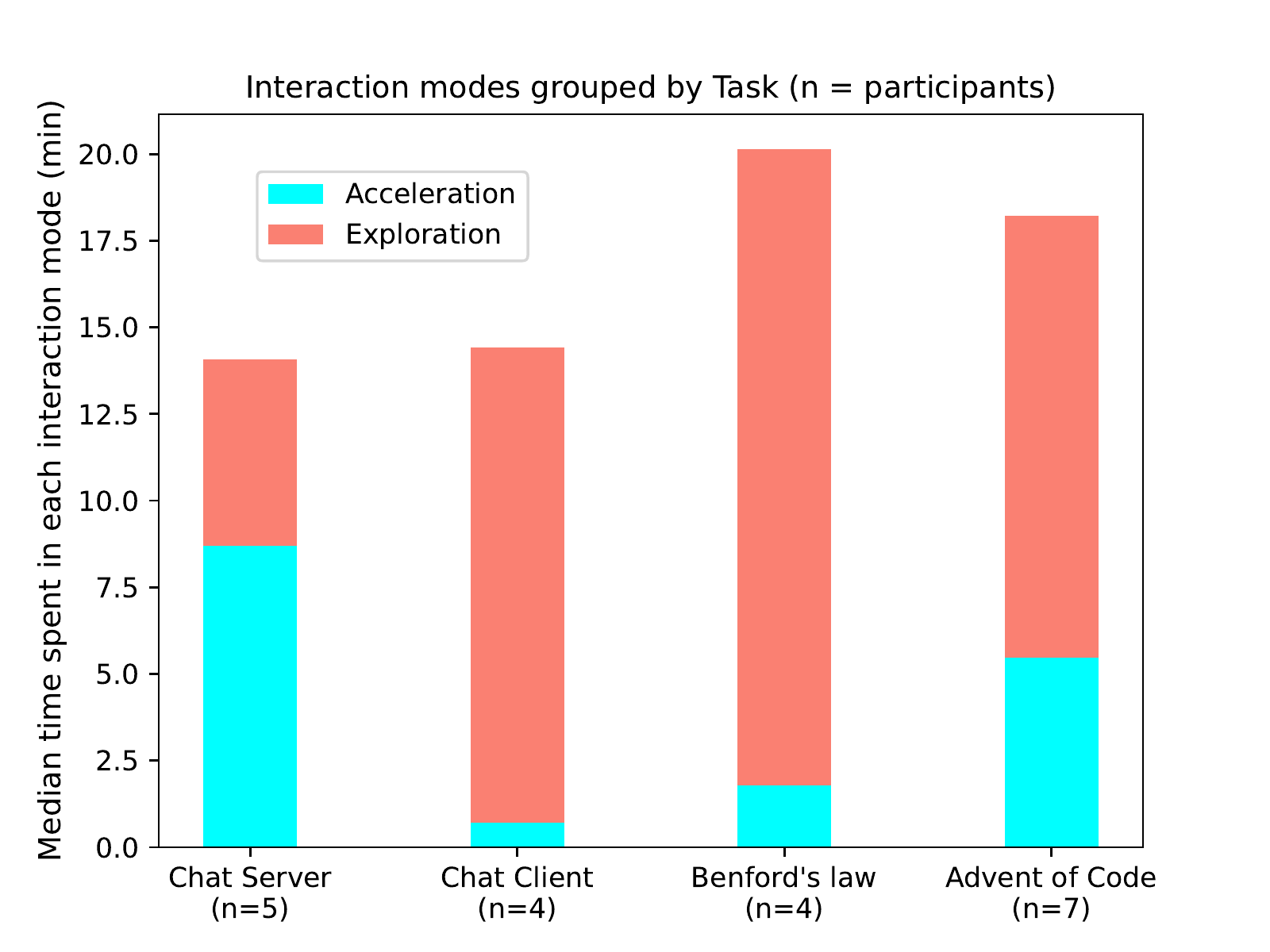}
		\captionsetup{width=.9\linewidth}
		\caption{Median time spent in acceleration vs exploration mode, grouped by task.}
		\label{fig:modes_task}
	\end{minipage}%
	\begin{minipage}{0.5\textwidth}
		\centering
		\includegraphics[width=\textwidth]{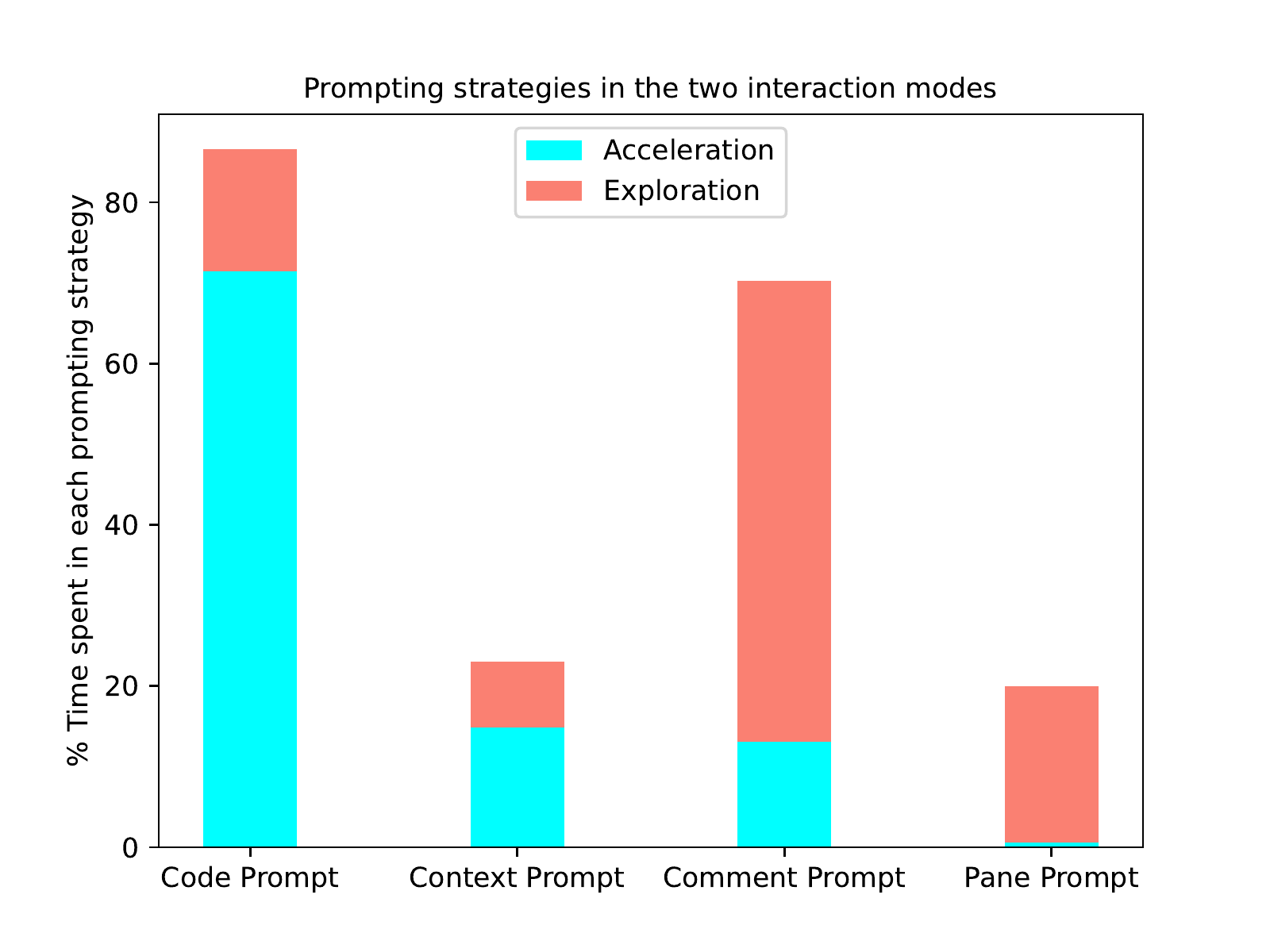}
		\captionsetup{width=.9\linewidth}
		\caption{Prevalence of prompting strategy as percentage of total prompting time.}
		\label{subtbl:the-table}
	\end{minipage}%
\end{figure}

\mypara{Nature of Task}
%
%
\autoref{fig:modes_task} shows the median time spent in each mode grouped by task. 
Both Chat Client and Benford's Law prominently feature interaction with unfamiliar APIs;
as a result, 
all participants in these two tasks spent considerably more time in exploration, 
irrespective of other varying factors such as language expertise and prior Copilot usage.
Advent of Code was more algorithmically challenging than the other tasks,
and also involved the File I/O API, which was somewhat unfamiliar to participants.
Both of these factors pushed participants to explore 
but there was more variance in the data than in Chat Client and Benford's Law:
for example, P16, who figured out the algorithm early on, 
spent more time accelerating (15.8 minutes) than exploring (3.4 minutes).
Chat Server, on the other hand, involved simple business logic, 
so participants leaned towards acceleration in this task.

\mypara{Programming Language}
We did not identify any noticeable differences in either total interaction time 
or ratio of acceleration to exploration between Python and Rust.
For the other two languages (Haskell and Java), we have too few data points to make any conclusions.
%

\mysubsubsection{Prompting strategies across interaction modes}
\label{sec:quant:promptByMode}

Our codebook identifies four strategies participants use to invoke Copilot:
\emph{code prompts}, \emph{context prompts}, \emph{comment prompts}, and the \emph{multi-suggestions pane}.
We can cluster the four prompting strategies into two categories: 
unintentional prompting (\autoref{sec:copilot:autocomplete}) and intentional prompting (\autoref{sec:copilot:exploration}).
Unintentional prompting involves participants invoking Copilot without explicitly meaning to.
For example, with \emph{code prompts}, the participant is often simply writing code when Copilot pops up a suggestion to complete their partially written line of code.
\emph{Context prompts} are those where Copilot generates suggestions even when the participant is not actively writing code. 
%
From the language model perspective, these two kinds of prompts are indistinguishable 
but we consider them distinct from the user interaction perspective.
Intentional prompting involves explicit intent from the participant.
This can be in the form of writing a natural-language comment intended for Copilot (\autoref{sec:exploration:comments}) 
or invoking the multi-suggestions pane by pressing \texttt{<ctrl> + <enter>} (\autoref{sec:exploration:multipleSuggestions}).

\autoref{subtbl:the-table} shows the aggregate percentage of times the 20
participants invoked Copilot using the four different prompting strategies.
We notice that in acceleration mode, 
the most commonly used prompting strategy is code prompts (71.4\%), 
with the other unintentional strategy, context prompts, coming in second (15.2\%).
%
The multi-suggestions pane is rarely used, 
which is consistent with our theory, 
since it would break the participant's flow.
In exploration mode, participants intentionally prompt with comments a lot more than in acceleration mode (57.2\% vs 13.1\%). 
%
The percentage of multi-suggestion pane prompts also shoots up in exploration mode as it
provides a rich body of suggestions for participants to explore from.

\mysubsubsection{Validation strategies across interaction modes and tasks}\label{sec:5.2}

\begin{figure}
	\centering
	\begin{subfigure}{.5\textwidth}
		\centering
		\includegraphics[width=\textwidth]{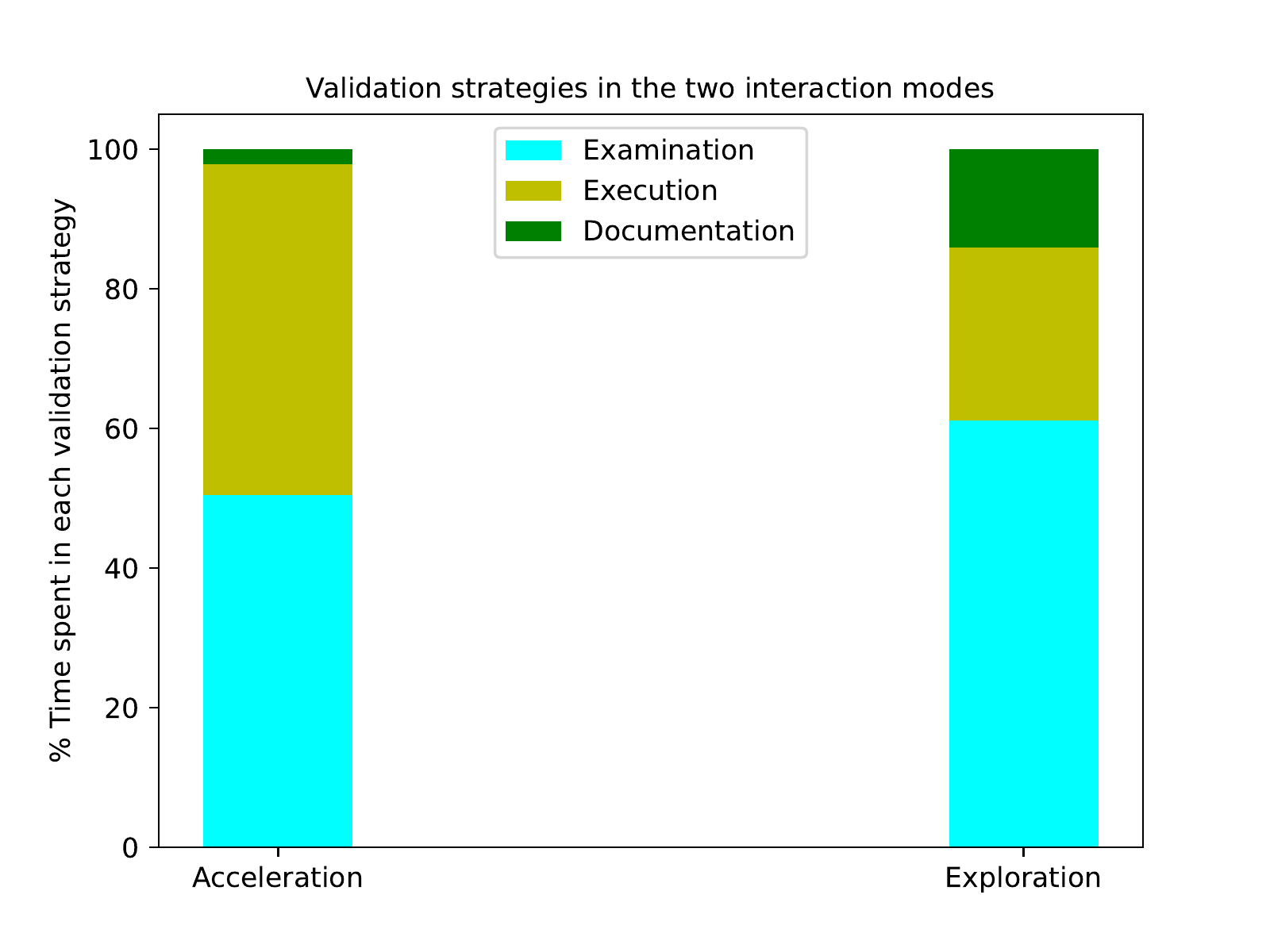}
		\caption{Aggregate time, split by interaction mode.}
		\captionsetup{width=\linewidth}
		\label{fig:val_modes}
	\end{subfigure}%
	\begin{subfigure}{.5\textwidth}
		\centering
		\includegraphics[width=\textwidth]{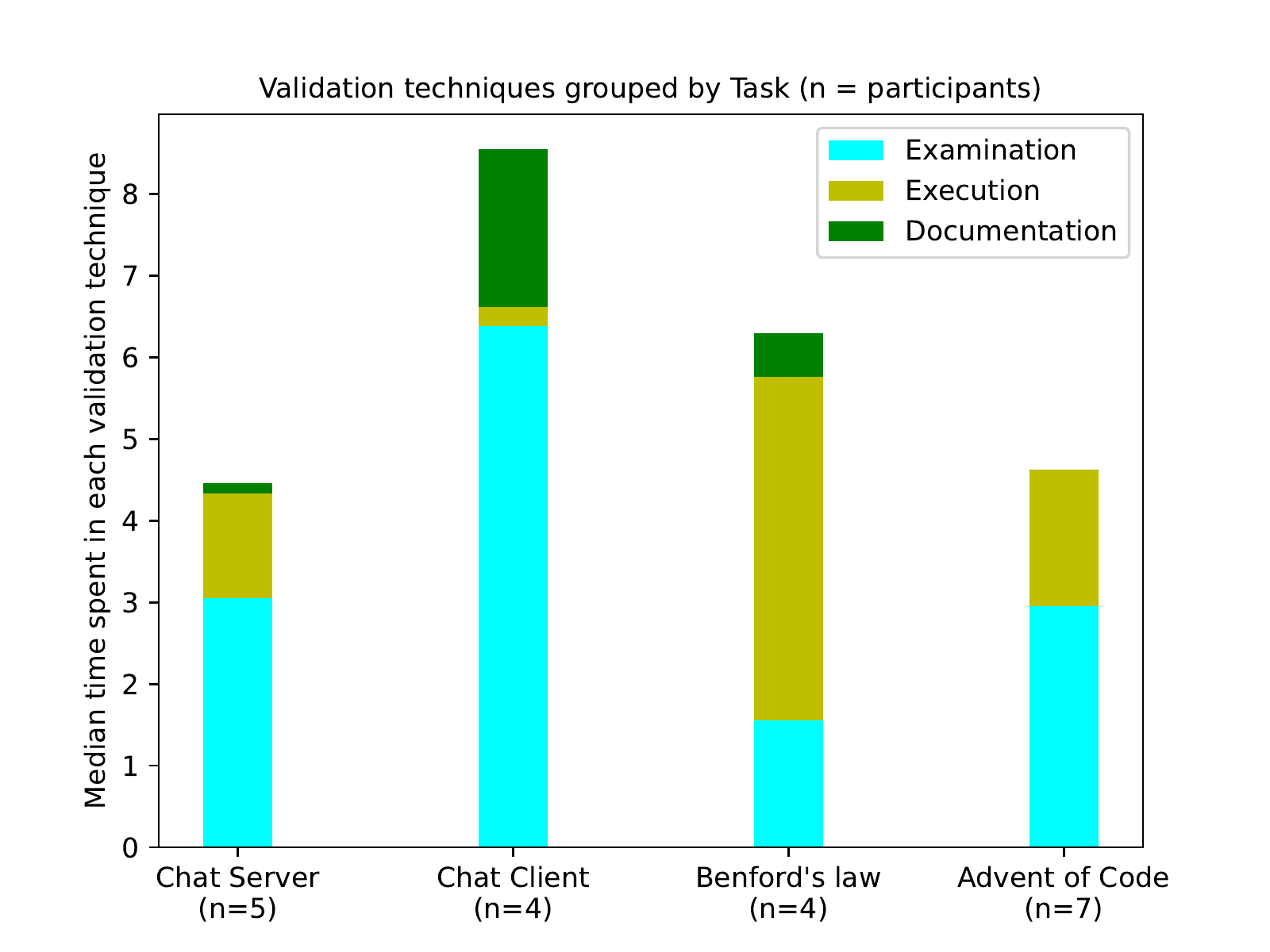}
		\captionsetup{width=.9\linewidth}
		\caption{Median time, grouped by task.}
		\label{fig:val_task}
	\end{subfigure}%
	\caption{Time spent in different validation strategies.}
	\label{fig:task_groups}
\end{figure}

Recall that \autoref{sec:exploration:validation} identified four different validations strategies:
\emph{examination}, \emph{execution}, \emph{static analysis}, and consulting the \emph{documentation}.
We measured the time participants spent in each of these strategies,
with the exception of static analysis,
which runs automatically in the background,
so it was hard for us to determine precisely when a participant was ``using'' its results.


\autoref{fig:val_modes} shows the percentage of validation time spent in each strategy, split by interaction mode.
Predictably, participants spent more time reading \emph{documentation} in exploration mode than in acceleration mode,
likely because exploration was commonly used when interfacing with unfamiliar APIs.
A somewhat surprising result is that \emph{execution} seems to be more prevalent during acceleration.
One reason for this is that during exploration the code is often incomplete and cannot be executed.
Another reason is simply that the remaining strategy, \emph{examination},
takes up significantly more time in absolute terms during exploration,
as participants carefully examine the code line by line as opposed to making quick decisions via ``pattern matching''.
We conclude that in exploration mode, 
programmers use validation strategies that \emph{aid comprehension} (careful examination, reading documentation),
while in acceleration more, they focus on strategies that provide \emph{rapid feedback} on code correctness (execution).

%

The nature of the task also has impact on the validation strategies,
as shown in \autoref{fig:val_task}.
The prevalence of complex and unfamiliar APIs in Chat Client
both increases the overall validation time for this task
and favors exploratory validation, such as examination and documentation.
Interestingly, the task with most time spent in execution is Benford's Law,
and not Advent of Code, which was explicitly designed to be easy to test
(it came with a test case).
We conjecture that Benford's Law was executed so often because it has visual output,
which is easy and exciting for programmers to inspect.


\subsection{Qualitative Analysis of LiveStreams}

We gathered additional evidence  in the form of five livestream videos to support our theory.
We present our findings from a qualitative analysis of these videos in this section.

\mysubsubsection{Data Collection}
The livestream videos were taken from Youtube (S1, S2, S4) and Twitch (S3, S5), and involved a developer using Copilot while constantly talking aloud to an audience.
%
S1 and S2 had Copilot turned on to solve Advent of Code tasks in Haskell and C\# respectively.
S3, S4 and S5 all did web-based programming tasks using Copilot in Javascript, Typescript, HTML, SCSS, and other web languages.
For example, S4's task was to build a Go game in Angular. 
While S1, S2 and S4 had well-defined tasks, S3 and S5 used Copilot for
exploratory tasks, in fact, S3 even asked their viewers to suggest random programming
tasks for Copilot.

\mysubsubsection{Qualitative Data Analysis}

One of the authors coded all the livestream videos
with the same closed codebook used to re-code our participant videos in
\autoref{sec:quant:analysis}.
We present the results from our qualitative analysis
and draw parallels to our bimodal theory of
acceleration and exploration.

\mysubsubsection{Acceleration Mode}
We observed that when the task was relatively well-defined (S1, S4, S5), acceleration mode was prevalent, consistent with our theory.
All streamers used Copilot for end-of-line completions in acceleration mode at
least once, accompanied with comments like,
``Yeah Copilot knows what I'm trying to do!''
In fact, S4 used Copilot only for end-of-line completions and said, ``I need to let the AI help more, I'm doing too much stuff myself.''
Streamers would only focus on small logical units, for instance, S2
accepted a long suggestion only to retain the structure of a for loop and the condition within.
S2 repeated this behavior when they just wanted to fill in the parameters of a function so they ended up deleting everything in a suggestion except the parameters.
S1 often used end-of-line completions to complete type signatures in Haskell, which would correspond to a logical unit.
As observed in our theory, the streamers would reject long suggestions that broke their flow (S1, S2, S4).
S4 exclaimed, ``Thank you, that's not what I want'' when Copilot suggested an
extremely long snippet while they were accelerating.
In addition, both S1 and S4 made only minor edits to suggestions accepted in acceleration mode, 
whereas S1 made relatively major edits to suggestions in exploration.

\mysubsubsection{Exploration Mode}

S3 and S5, who worked on exploratory tasks, spent considerably more time in exploration mode than in acceleration.
Streamers were willing to write a lot of comments while in exploration mode (S2,
S3, S5).
S5 tried to use Copilot to generate documentation and said, ``as a person who usually writes comments after writing code, Copilot might change the way I code''.
S3 had an interesting way of prompting Copilot: by writing unusually descriptive function names instead of comments.
S3 and S5 often used the multi-suggestion pane as a fallback option when the inline suggestions did not meet their expectations.
S3 expected the multiple suggestions to be diverse and was sometimes disappointed when they were not.
In addition to using the pane, S5 also explored multiple suggestions inline by pressing tab. We did not observe this behavior in our main study, because neither we nor our participants were aware of this feature.

\mysubsubsection{Validation Strategies}

We observed the same validation behavior as seen in our theory in all the livestream videos.
After accepting a suggestion, S3 said, ``Let's just check if this part works'' and S1 echoed, ``I think Copilot wrote that for me, let me just check''.
S1 and S2 constantly validated their code using the test inputs provided by the Advent of Code tasks and also used specific test inputs for debugging code.
All streamers spent considerable time in code examination as a validation strategy both inline (S1, S2, S4, S5) and in the multi-suggestion pane (S3, S5).
S2 and S3 referred to API documentation using web search to validate Copilot's code while S4 resorted to reading in-IDE API documentation as a form of validation.
S3 and S4 whose tasks involved building a website ran their webpage remotely as a validation strategy.

The blame game of who wrote the buggy code was also observed in the livestreams.
While debugging their code, S5 expressed this by saying, ``not sure if they are my bugs or Copilot's bugs.''
S3 had a bug that they were baffled by, turns out it was some residual code from Copilot's suggestion which they forgot to delete.
S5 summed up Copilot's behavior as being a ``mixed bag, when it understands what I want it feels like it's reading my mind. Otherwise it produces random code.''
Streamers were generally confident using Copilot for writing boilerplate, repetitive code (S3, S4).

\section{Recommendations}\label{sec:recs}

This section outlines recommendations for how programming assistants
could be improved in the future,
%
We classify these suggestions into two categories:
improving the way programmers could provide \emph{input} to a future tool,
and improving the kinds of \emph{output} the tool could generate.

\subsection{Better Input}

\mypara{Control over the context}
There was general confusion among participants about how Copilot uses their
code to provide suggestions.
Some participants were unsure how much code Copilot can take into context,
for example, P8 theorized a hard limit to the input length:
``I think the README is too long and complicated for it to actually
extract [helpful information]''.
Other participants (P8, P10, P15, P18) mentioned they were unsure about which
pieces of
information Copilot had extracted about their local codebase.
Specifically, there appeared to be a broad misconception that commenting out code made it invisible to Copilot,
despite those same participants using comment prompts.
P20 ``assumed it wouldn't be aware of code if [they] commented it out''.
We also observed participants (P2, P3, P4 P6) comment out code generated by Copilot
in an attempt to get it to generate an alternative suggestion.

Participants that \emph{were} aware of Copilot's sensitivity to context
wanted to have more control over that context.
Some participants wanted to give Copilot \emph{specific context}:
in describing their work outside of the study,
P15 mentioned poor suggestions from Copilot and wished they could emphasize a
subset of their code
(\ie niche libraries they imported),
so they could feel more confident that the suggestions were relevant to their
code.
Others, P4 and P12, wished to query Copilot
with a natural-language prompt
\emph{without} any code context,
just as they would query StackOverflow.
%
%

In order to achieve this control,
participants wanted Copilot to provide dedicated \emph{syntax}.
For example, P2 wanted Copilot to use a specific function,
and tried to achieve this by ``using the function name in backquotes''.
P18 asked:
``Is there a way to prompt Copilot into suggesting a data structure?''
Finally, P4, when looking for examples of using the \texttt{del} operation in Python,
wanted to explicitly ask Copilot to show only ``syntax examples''.

%
%
%
%


Based on these observations,
future tools could give programmers ways to customize the context.
For example, a future tool could provide a scratchpad to isolate general,
StackOverflow-style prompts from the rest of the codebase.
It could also provide expert prompt syntax,
similar to advanced operators in Google search;
for example, including \texttt{:use plt.show()} in a comment prompt
might restrict the assistant's suggestions to only those snippets using the
expression \texttt{plt.show()}, like the work of~\cite{notOnlyByExample}.
Finally, programmers would likely appreciate a separate type of comments
that make code invisible to the tool.

\mypara{Cross-language translation}
P13 said that they were more familiar with Julia
than the task language (Python),
and at some point they wrote some Julia code which Copilot then translated to Python.
%
This type of interaction opens up the possibility of users giving
prompts in programming languages they are more familiar with.
The task for Copilot then becomes a cross-language translation
task.
It would be interesting to fine-tune Copilot for this particular task, by
training it on equivalence classes of syntactic constructs in different
programming languages.

\subsection{Better Output}

\mypara{Awareness of the interaction mode}
Perhaps the most important outcome of our study
is the bimodal nature of programmers' interactions with Copilot:
they are either in an acceleration or exploration mode.
We conjecture that the user experience could be improved
if the tool were aware of the current interaction mode and adjusted its behavior accordingly.
In acceleration mode, it should not break the programmer's flow
(P6 mentioned that they intentionally turned Copilot off
because it disrupted their workflow).
To this end, the tool should avoid low-confidence suggestions---%
which are unlikely to be accepted---%
and long suggestions---%
which distract the programmer.

Going beyond simply avoiding multi-line suggestions,
the tool could be made more aware of how the code is divided into logical units.
As we mentioned in \autoref{sec:acceleration:logicalUnits},
programmers in acceleration mode focus on a single logical unit of code at a time,
which is often one line,
but can also be shorter (the next function call in Haskell)
or longer (an entire loop).
It would be interesting to explore if we can make the scope of Copilot's suggestions
match the scope the programmer's current focus.
Participants also mentioned that it would be helpful
if Copilot gave suggestions more selectively as opposed to being always on.
This could be achieved, \eg, by reinforcement learning
to obtain a policy for when Copilot should intervene,
based on the local context and programmer's actions.

\mypara{Exploring multiple suggestions}
As we mentioned in \autoref{sec:exploration:multipleSuggestions},
in exploratory searches, programmers commonly used the multi-suggestion pane,
but also often got overwhelmed by the results they saw there.
Several participants had trouble identifying meaningful differences between the suggestions
(P1, P4, P6, P7, P13).
This observation motivates the need for a tool
that would help programmers explore a large space of suggestions,
perhaps similarly to how Overcode~\cite{overcode} supports exploring
a space of student solutions to a programming assignment.

\mypara{Suggestions with holes}
Recall from \autoref{sec:exploration:acceptAndEdit}, that when
programmers modify suggestions, they often keep control-flow features and
little else, as seen for P1, P2, P8, P11, P17, and P18.
%
%
Based on this observation,
programmers would likely benefit from \emph{suggestions with holes},
where the tool only generates control structures,
which users are likely to understand quickly,
leaving their bodies for the programmer to fill out
(either by hand, or by giving more targeted prompts to the tool).
For example, P2 explicitly mentioned that
``if [Copilot] gives me a mostly filled out skeleton, I can be the one who fills out holes''.
Recent work by~\cite{guo2021learning} generated holes in their suggestions where
the underlying model
had low confidence.
%

Low-confidence suggestions are not the only motivation for a hole: participants
reported feeling frustrated and distracted by large code snippets.
When offered these large snippets, some participants felt Copilot was
forcing them to jump in to write code before coming up with a
high-level architectural design. P4 said:
\begin{myquote}
``I wrote code as one might read code, rather than the way I might write it
	which is generally top-down, where I will fill in the control structure and
	then I'll do the little bits and pieces after I build in the full control
	structure. It made me jump in to write code instead of the normal way.''
\end{myquote}
P16 normally writes a high-level design first and then
gets to function implementations---as the grounded theory
from~\cite{Lubin_Chasins_2021} describes of functional programmers.
Other participants (P2, P3, P4, P5, P7, P8) also felt Copilot forced significant change on
their code authorship process.
%
Based on our observations, future tools should mind how large code blocks can
break the user's natural development flow, instead offering code holes for users
to fill in when ready.

%
%

\mypara{Always-on validation}
Several participants (P2, P14, P16) wished to have better tool support for validating suggestions.
For example, P16 wanted to set up property-based testing~\cite{Claessen_Hughes_2000}
to run automatically on Copilot suggestions.
%
P14 wished they had \emph{projection boxes}~\cite{Lerner_2020},
a live programming environment that constantly displays runtime values of relevant variables
(usually on a single test input).
In the future, IDEs could couple code-generating models
with some kind of always-on validation,
in order to make the process of evaluating code suggestions
less taxing for the developer.

\section{Related Work}\label{sec:related}

\mypara{Usability of Copilot}
The closest to our work is the study by \citet{Vaithilingam_2022},
which also evaluates Copilot.
%
The main differences are:
\begin{enumerate*}[(1)]
\item their study is on stand-alone tasks, whereas ours includes tasks that require contributing to an existing codebase;
\item their study is comparative and focuses on the rate and time of task completion with and without Copilot's help;
\item their study only used Python, whereas we used several programming languages.
\end{enumerate*}
In our study, we explicitly stepped away from the common comparative setting,
where participants are given well-defined stand-alone tasks,
and the goal is to collect quantitative data on how well and quickly they complete the tasks,
with and without the tool under evaluation.
Instead, we chose more open-ended tasks in the context of an existing codebase,
which we believe is closer to the real-world use case.
Further, instead of skewing quantitative answers to predefined research questions,
we chose the grounded theory approach,
with the general goal of finding patters in programmers' behavior when they interact with Copilot;
we believe this approach is complementary to the quantitative studies.
Finally, our usage of multiple languages enables inter-language comparisons and more generalizable conclusions.

On the other hand, \citet{Vaithilingam_2022} also report several qualitative findings.
Most of them agree with ours, such as:
that Copilot often provides a good starting point for programmers who do not know how to approach the task,
that programmers are generally willing to repair code suggestions,
but Copilot-generated code is harder to debug.
There are also some differences; fore example,
half of their participants (12/24) said they had trouble understanding and modifying Copilot-generated code,
whereas our participants did not seem to share this difficulty;
this might be because our study is with more experiences developers:
only one participant in our study was an undergraduate student, whereas 10/24 in their study were undergraduates.

\mypara{Usability of other LLM tools}
Beyond Copilot, \citet{jiang2022discovering} conducted a user study to
analyse the interaction of developers with a natural language to code tool called GenLine.
GenLine is similar to Copilot but involves explicitly invoking a command within a text editor.
Similar to our findings, developers in their study were willing to
rewrite the natural language prompt to clarify their intent
and expressed
the need for a syntax to communicate with the model more clearly.
However, their findings were mainly centered around prompting strategies whereas we did a
more comprehensive analysis of developer interactions with Copilot.
Moreover, the tool was not integrated in the participant's daily workflow like in our study.
In a similar vein, \citet{Xu_2021} investigated the usefulness of an NL-to-code plugin
previously developed by the same authors \cite{Xu_2020}.
They found no statistically significant difference in task completion times or correctness scores
when using the plugin,
and the participants' feedback about the plugin was neutral to slightly positive.
We conjecture, however, that these findings are not as relevant anymore,
thanks to recent breakthroughs in large language models,
which significantly increased the quality of generated code.
In another related study, \citet{Weisz_2021} interviewed IBM software engineers
about their experience with a neural machine translation tool
for translating code between programming languages.
This study focuses on the engineers' code validation strategies
and future UI features that might help with this task,
such as confidence highlighting and alternative translations;
in this sense, their study is complementary to our work,
conducted in the context of a different task (language-to-language translation).

\cite{sarkar2022like} compiled observations from the above user studies and additionally gathered experience reports of programming assistants usage from Hacker News.
The compiled observations were similar to what we found---prompting is hard, validation is important, and programmers use assistants for boilerplate, reusable code.
%
%
There are a few other industrial-grade programming assistants powered by statistical models,
such as TabNine~\cite{tabnine} and Kite~\cite{kite},
but we are not aware of any research on their usability.


\mypara{Usability of program synthesis tools}
Another approach to code generation,
is the more traditional, search-based program synthesis.
As program synthesis technology matures,
it becomes increasingly common to evaluate the usability of synthesizers on human subjects.
Many of these usability studies are for domain-specific synthesizers
targeting API navigation~\cite{hplus},
regular expressions~\cite{regae,Zhang_2021},
web scraping~\cite{rousillon},
or data querying, wrangling, and visualization~\cite{wrex,falx,Zhou_2022}.
These studies usually focus on measuring the tool's effect on task completion rates and times,
which is less relevant to our questions.
%
%
The work on RESL~\cite{resl} and Snippy~\cite{snippy,loopy}
include user studies of general-purpose programming-by-example synthesizers for JavaScript and Python.
Although both also focus mainly on task completion times,
they do make some interesting qualitative observations.
%
For example, \citet{snippy} observe that one of the main barriers to the usefulness of the synthesizer
is the so-called \emph{user-synthesizer gap},
\ie the programmer's overestimation of the synthesizer's capabilities;
we observed a similar phenomenon in our study (see \autoref{sec:exploration:trust}),
although it appears to be less prominent in LLM-based tools,
since their performance degrades more gradually with the complexity of the task.

\cite{Jayagopal_Lubin_Chasins_2022} study how undergraduate students
learned to use six different synthesizers---Copilot among them---with
different interaction modes.
%
Not all of the themes they identify are applicable to Copilot,
but those that are, are corroborated and explored in more depth in our study.
For example, they identify that novice participants would often accept then
modify code.
We support and extend this (\autoref{sec:exploration:acceptAndEdit}),
adding that this is characteristic of exploration mode, 
which indeed occurs more commonly in novices.




\mypara{Grounded Theory for software development}
Grounded Theory (GT) has a relatively long history in software-related fields,
with its application to software engineering dating as far back as 2004~\cite{Carver_2004}.
\cite{Stol_Ralph_Fitzgerald_2016}~provide a survey and a critical evaluation
of 93 GT studies in software engineering.
Recently, GT has also drawn interest in the programming languages community:
\citet{Lubin_Chasins_2021} study how statically-typed functional programmers write code,
and deliver a set of guidelines meant for functional language tool-builders.





\section{Limitations and Threats to Validity}
%
%
%
Our participants worked on tasks of our design, 
as opposed to their own projects.
If they were working in a more familiar codebase and without the time pressure
of a study, their interactions could have been different.
%
Moreover, our tasks focused only on code authorship,
as opposed to refactoring, testing, debugging,
or other common aspects of software engineering.
We consider these beyond the scope of this study,
although our participants did occasionally get a chance to test or debug their code.
%

We recruited 20 participants, with a skew towards those in academia, hardly a
representative sample of all programmers.
%
Similarly, although we tried to diversify the type of tasks our participants were solving
and the programming languages they were using, 
other kinds of tasks and languages could have lead to different interactions.
%

%
%
11 of our participants had not used Copilot before the study,
and hence might not be representative of regular users of the tool.
We gave all participants a 5-minute training task so they could familiarize themselves with Copilot,
and yet we observed that first-time users were sometimes over-reliant on Copilot,
in a way that prior users were not. 
%
We chose to include new users in our study since the majority of programmers
in the wild have never used a code-generating model.
%
Meanwhile, our participants who already had access to the tool may have formed a
usage pattern (or dis-uage pattern in the case of P6) based on poor experience
early in the technical preview,
where its behavior may have been rapidly changing.
Ideally, we would have liked to observe programmer over a longer period of time,
in order to study how their usage patterns changed over time,
but this was not feasible given the time constraints of the study.

Finally, the research on code-generating models is progressing very rapidly,
and it is possible that new technological breakthroughs will soon render our findings obsolete.
That would be a nice problem to have indeed!

\begin{acks}
The authors would like to thank Devon Rifkin from GitHub for his assistance
with getting Copilot access for our study participants.
This work was supported by NSF grants 2107397 and 1955457.
\end{acks}

   \nocite{*}
	\bibliography{main}
	\clearpage
	\appendix

\end{document}